\documentclass{ptephy_v1}

\preprintnumber{XXXX-XXXX} 

\usepackage{bm} 
\usepackage{url}
\usepackage{array}
\usepackage{fancybox}
\usepackage{multirow}
\usepackage{appendix}
\usepackage{epsfig}
\usepackage{cases}

\newcommand{\newc}{\newcommand}
\newcommand*{\ba}{\begin{eqnarray}}
\newcommand*{\ea}{\end{eqnarray}}
\newcommand*{\bb}{\begin{framed}}
\newcommand*{\eb}{\end{framed}}
\newc{\be}{\begin{equation}}
\newc{\ee}{\end{equation}}
\newc{\bea}{\begin{eqnarray*}}
\newc{\eea}{\end{eqnarray*}}

\newcommand{\simgt}{\lower.5ex\hbox{$\; \buildrel > \over \sim \;$}}
\newcommand{\simlt}{\lower.5ex\hbox{$\; \buildrel < \over \sim \;$}}

\newcommand*{\rd}{{\rm d}}
\newcommand*{\rE}{{\rm E}}

\newcommand*{\mpl}{M_{\rm pl}}

\def\({\biggl(}
\def\){\biggr)}
\def\[{\biggl[}
\def\]{\biggr]}

\newcommand{\equref}[1]{Eq.~(\ref{eq:#1})}
\newcommand{\figref}[1]{\figref{fig:#1}}

\newcommand{\apj}{Astrophys. J. }
\newcommand{\apjs}{Astrophys. J. Suppl. }
\newcommand{\apjl}{Astrophys. J. Lett. }

\newcommand{\physrep}{Phys. Rep. }
\newcommand{\ptp}{Prog. Theor. Phys. }

\newcommand{\ptep}{Prog. Theor. Exp. Phys. }

\newcommand{\aap}{Astron. Astrophys. }

\newcommand{\nat}{Nature }

\newcommand{\prc}{Phys. Rev. C}
\newcommand{\prd}{Phys. Rev. D}
\newcommand{\prl}{Phys. Rev. Lett. }

\newcommand{\araa}{Ann. Rev. Astron. Astrophy. }
\newcommand{\mnras}{Mon. Not. Roy. Astron. Soc. }

\newcommand{\nphysa}{Nucl. Phys. A}

\newcommand{\jcap}{JCAP}

\begin{document}

\title{Neutron star cooling in modified gravity theories}

\author[1,*]{Akira Dohi}
\affil[1]{Department of Physics, Kyushu University,  744 Motooka, Nishi-Ku, Fukuoka Fukuoka 819-0395, Japan\email{dohi@phys.kyushu-u.ac.jp}}
\author[2]{Ryotaro Kase,}
\affil[2]{Department of Physics, Faculty of Science, Tokyo University of Science, 1-3, Kagurazaka, Shinjuku-ku, Tokyo 1}
\author[3]{Rampei Kimura,}
\affil[3]{Waseda Institute for Advanced Study, Waseda University
19th building, 
1-21-1 Nishiwaseda, Shinjuku-ku, Tokyo 169-0051, Japan}
\author[1]{Kazuhiro Yamamoto,}
\author[1]{Masa-aki Hashimoto}



\begin{abstract}
We study thermal evolution of isolated neutron stars in scalar-tensor theories for the first time. Whether the rapid cooling due to the direct Urca process occurs or not is an interesting question in the viewpoint of the temperature observation of isolated neutron stars. Moreover, investigation of the cooling effect of nucleon superfluidity also has the large uncertainties though it is important in modern cooling theory. The cooling effect is typically influenced by the proton fraction and the central density. If a fifth force is mediated due to modification of gravity, the relation between the central density and mass of neutron stars differs from one in general relativity, and the cooling curve is also naively expected to be varied. We find that an unscreened fifth force near the surface of neutron stars changes mass-central density relation, and the direct Urca process can be triggered even for neutron stars with smaller mass. We also present
cooling curves including nucleon superfluidity under the scalar-tensor theory. These results show that it might be useful to test gravitational theories with cooling observations of neutron stars.
\end{abstract}

\subjectindex{E03 Alternative theory of gravity, E25 Stellar structure and evolution, E32 Neutron stars}
\maketitle

\setcounter{footnote}{1}
\section{Introduction}

The recent observations of gravitational waves from binary system \cite{Abbott:2016blz} play an important role in astrophysics and cosmology for testing the relativistic nature of gravity. One of the most significant results of LIGO and VIRGO is the detection of gravitational wave event (GW170817) from a neutron star (NS) binary and its electromagnetic counterpart (GRB170817A) \cite{TheLIGOScientific:2017qsa}, which provides a stringent constraint on the propagation speed of the gravitational waves \cite{Monitor:2017mdv}. This observation enables us to exclude a large theoretical space of gravitational theories beyond general relativity, which could be responsible for explaining the accelerated expansion of the universe \cite{Lombriser:2015sxa,Lombriser:2016yzn,Ezquiaga:2017ekz,Creminelli:2017sry,Baker:2017hug,Sakstein:2017xjx}. Among a vast class of models proposed so far, the remaining scalar-tensor theories in the Horndeski theory \cite{Horndeski:1974wa} after imposing this constraint consist of a  non-minimal coupling to gravity with the kinetic gravity braiding (K-essence plus cubic galileon) \cite{Deffayet:2010qz}, including $f(R)$ theories \cite{Carroll:2003wy,Hu:2007nk,Starobinsky:2007hu}. In such theories, the modification of gravity can be significant at cosmological scales in general, and hence observations such as type Ia supernovae \cite{Riess:1998cb,Perlmutter:1998np}, cosmic microwave background \cite{Ade:2015xua}, large scale structures \cite{Gil-Marin:2014sta,More:2014uva,KY2010} and clusters of galaxies \cite{terukina1,terukina2} are able to pin down parameter space in modified gravity theories.
In addition, the solar-system experiments allow us to test gravity at short distances even for extremely small deviations from general relativity, and it is therefore possible to put a stringent constraint on a certain class of modified gravity theories \cite{Will:2014kxa}.
During the last few decades, these tests are well-formulated in a various context of observations for  modified gravity theories, and the future observations are expected to improve these observational constraints (See for review e.g., \cite{Koyama:2015vza}). These cosmological observations and solar-system experiments can, however, test gravity only at low energy scales; therefore, testing strong gravity regime could provide extra information of modified gravity theories, and non-trivial effects from relativistic objects, especially NSs, should be extremely important in future observations, and such tests in strong gravity environments could be a powerful tool for measuring deviation from general relativity.

NSs, which are born after supernova explosion, cool down by emitting a large number of neutrinos and photons. In the absence of companion stars, isolated neutron stars (INS) cool by the neutrino emission for the time $t\lesssim 10^{5}~{\rm yr}$ after their birth, since then the photon-emission cooling follows. In the era of the neutrino cooling, thermal evolution of INS depends on the interior physics of the equation of state (EOS) including nucleon superfluidity. The most important problem for INS cooling is whether the direct Urca (DU) process, which is the strong neutrino emission through $\beta$ and inverse $\beta$ decay, occurs or not. The DU process is unnecessary for most of the observed neutron stars, which can be explained by the slow cooling processes with an additional cooling effect due to nucleon superfluidity, {\it minimal cooling scenario~\cite{Page2004}. The minimal cooling scenario is therefore widely recognized to be the most successful standard cooling theory for most of temperature observations.} However, some of neutron stars are thought to have enhanced cooling such as the DU process in order to explain their observed low temperature. For example, the pulsar J0205+6449 in supernova remnant 3C58 and RX J0007.0+7302 are cold enough to need the fast cooling processes~\cite{Page2004}. For accreting neutron stars in steady state, SAX J1808.4$-$3658 and 1H 1905+000 have low quiescent luminosity for high mass accretion rate~\cite{Heinke2009} and therefore they are thought to be the strong evidence of occurrence of fast cooling processes. Moreover, some cold compact objects such as, G127.1+0.5, G084.2+0.8, G074.0$-$8.5, G065.3+5.7~\cite{Kaplan2004}, and G043.3$-$0.2~\cite{Kaplan2006}, have been detected though it's unknown whether they are NSs or black holes, respectively. This implies that, if any one of them is NS, a fast cooling process should occur in this star~(e.g., see Figure.~11 in Ref.~\cite{Beznogov2020}). Focusing on the occurrence of the DU process, the thermal evolution of neutron stars is determined by the $Y_p$--density relation~\cite{Lattimer1991a} and the mass as well as the EOS. Therefore, the EOS could be constrained by observations of thermal evolution of INSs~(e.g., \cite{Dohi2019}). The recent observations indeed have rejected some of the EOSs~\cite{Demorest2010,Steiner2010,Antoniadis2013,Abbott2018,Cromartie2019,Riley2019,Miller2019}, though there are uncertainties in the mass-radius relation as well as the possible effects of exotic particles. Moreover, the change of the total neutrino emissivities by nucleon superfluidity is unclear for especially triplet gap of neutrons because of uncertain nuclear potentials~\cite{Takatsuka1972,Amundsen1985b, Baldo1992,Takatsuka2004}.  Due to these uncertainties, there can be many scenarios to account for the observations of the age and surface temperature \cite{Page2004,Yakovlev2004}. 

The observations of NS structure are rapidly advancing. For example, a few heavy NSs with $M\sim2~M_{\odot}$ have been recently discovered by relativistic Shapiro delay combined with the mass function of Kepler motion~\cite{Demorest2010,Antoniadis2013,Cromartie2019}, where $M_{\odot}$ is the solar mass. Especially, the millisecond pulsar PSR J0740+6620 discovered using the Green Bank Telescope has the most heaviest mass so far, which is estimated to be $M=2.14^{+0.10}_{-0.09}~M_{\odot}$ within $1\sigma$ region~\cite{Cromartie2019}. Such constraints about massive NSs enable us to reject many soft EOSs. Furthermore, the NICER has reported the observations of another millisecond pulsar PSR J0030+0451 , where the observational mass $M$ and the equatorial radius $R_{\rm eq}$ are estimated to be $M=1.34^{+0.15}_{-0.16}~M_{\odot}, R_{\rm eq}=12.71^{+1.14}_{-1.19}~{\rm km}$~\cite{Riley2019} and $M=1.44^{+0.15}_{-0.14}~M_{\odot}, R_{\rm eq}=13.02^{+1.24}_{-1.06}~{\rm km} $~\cite{Miller2019} within $1\sigma$ regions, respectively.\footnote{Although the same observational data are used in the Refs.~\cite{Riley2019,Miller2019}, the discrepancy between these constraints stems from independent analysis of these ground.}

Going back to modified gravity theories, an extra degree of freedom in scalar-tensor theories typically modifies the law of gravity even for small scales, and therefore screening mechanisms such as the Chameleon mechanism \cite{Khoury:2003aq,Khoury:2003rn} or the Vainshtein mechanism \cite{Vainshtein:1972sx} are crucial to pass stringent tests in the solar system. On the other hand, a screening mechanism might not completely hide the fifth force in NSs as reported in \cite{Kase:2019dqc}. Once the mass-radius relation (or the mass-central density relation) are modified, one would expect that thermal evolution of NSs can be also different from general relativity. If this is true, there should be deviation in cooling curves compared with that of general relativity, and one might be able to test modified gravity theories by using cooling curves of NSs. 
With this in mind, for the first time, we investigate how the cooling curve of NSs changes in the scalar-tensor theory proposed in \cite{Kase:2019dqc}. Although this model can be tightly constrained by the local gravity experiments \cite{Burrage:2017qrf}, it could potentially change the structure of neutron stars from the one in general relativity as shown in \cite{Kase:2019dqc} and is therefore suitable to see the effect on the cooling curve. The structure of NSs in other modified gravity theories such as $f(R)$ and galileon theories has been intensively investigated in many literatures \cite{KM2008,Babichev,Canate:2017bao,Sultana:2018fkw,Cooney:2009rr,Arapoglu:2010rz,Orellana:2013gn,Astashenok:2013vza,Ganguly:2013taa,Yazadjiev:2014cza,Resco:2016upv,Cheoun2013,Kobayashi2018,Ogawa2020}.  In the present paper, we show how the attractive nature of the fifth force affects to the cooling curve, which can be applicable for other Chameleon theories.

This paper is organized as follows. In Sec.~II, we review the model and structure of NSs studied in \cite{Kase:2019dqc} and  the EOSs adopted in the present paper. In Sec.~III, we give an overview of NS cooling, including observations of isolated NSs. In Sec.~IV. we show  cooling curves of NSs in the scalar-tensor theory. 
Sec.~V is devoted to discussions and conclusion. 
Details on the relation between Jordan and Einstein frame are summarized in Appendix \ref{app:EF}. In Appendix \ref{app:chameleon}, we briefly review the Chameleon mechanism. In Appendix \ref{app:BD}, we present the cooling curves of NSs in the massless Brans-Dicke theory as a complimentary material.
Throughout this paper, we use the reduced Planck mass $\mpl=1/\sqrt{8\pi G}$,
the density 
$\tilde{\rho}_0
=m_n n_0=1.6749 \times 10^{14}~{\rm g} \cdot {\rm cm}^{-3}$ with $n_0=0.1~{\rm (fm)}^{-3}$,
where $m_n=1.6749 \times 10^{-24}$~g is the neutron mass, 
and the distance $r_0 
={c}/{\sqrt{G \tilde{\rho}_0}}=89.664~{\rm km}$ for normalization.

\section{Theory}

In this section, we give a brief summary of theoretical set-up, basic equations, and numerical solutions of NSs. 
We consider the following model in scalar-tensor theories \cite{Tsujikawa:2008uc},
\ba
 S = \int d^4 x \sqrt{-g} \Biggl[{\mpl^2 \over 2} F(\phi) R -{1 \over 2} (1-6Q^2) F(\phi) \partial_\mu \phi \partial^\mu \phi - V(\phi) \Biggr] + S_{\rm m} [g_{\mu\nu}, \Psi_{\rm m}],
 \label{action}
\ea
where $g$ is the determinant of metric tensor $g_{\mu\nu}$, $R$ is the Ricci scalar, 
$Q$ is a constant, $\Psi_{\rm m}$ describes the matter field minimally coupled with $g_{\mu\nu}$,
and the non-minimal coupling $F(\phi)$ is chosen to be
\ba
F(\phi) = e^{-2Q\phi /\mpl}\,. 
\label{coupling}
\ea
We focus on the model with the following potential 
\ba
V(\phi) &=&{1 \over 4}\lambda \phi^4 {\rm ~~~ with~~~} Q= - 1/\sqrt{6}\,.
\label{potentialV}
\ea
Here, the value of $Q$ in \eqref{potentialV} 
is chosen such that $\phi$ becomes a canonically normalized field in the Einstein frame 
without a filed redefinition \cite{Tsujikawa:2008uc}, 
and the infinite limit $\lambda \to \infty$ corresponds to general relativity. In this model, the canonical kinetic term for the scalar field is apparently absent due to the choice of $Q$, however, the scalar field regains a kinetic term from the non-minimal coupling with gravity, as we will see below. The structural features of the action \eqref{action} with $Q=-1/\sqrt{6}$ is the well-known relation with the metric $f(R)$ gravity \cite{DeFelice:2010aj} given by
\ba
 S = \int d^4 x \sqrt{-g} {\mpl^2 \over 2} f(R)
 + S_{\rm m} [g_{\mu\nu}, \Psi_{\rm m}]
 \,, \label{actfR}
\ea
 through an appropriate choice of the function $F$ and the potential $V$,
\ba
V(\phi) =  {\mpl^2 \over 2}  (FR-f), \qquad F(\phi)={\partial f \over \partial R} \,. 
\ea
The $\phi^4$ potential given in (\ref{potentialV}) arises from the $f(R)$ theories described by 
$f(R)=R+aR^{4/3}$ where $a$ is a constant \cite{Kase:2019dqc}.
Although the theories (\ref{action}) and (\ref{actfR}) are equivalent, for simplicity, we use the scalar-tensor action \eqref{action} throughout this paper. We keep the parameter $Q$ in this section for the sake of completeness, and the case in the massless Brans-Dicke theory is discussed in Appendix~\ref{app:BD}. 
The model with the $\phi^4$ potential potentially possesses a screening mechanism for the fifth force 
in dense environments
\cite{Khoury:2003aq,Khoury:2003rn}. Thanks to this screening effect called the Chameleon mechanism, the model
with the $\phi^4$ potential can be consistent with the local gravity experiments (see e.g., \cite{Burrage:2017qrf}). In Appendix \ref{app:chameleon}, we summarized a quick introduction to the Chameleon mechanism. 
In this paper, we choose relatively larger $\lambda$ where the corresponding Compton wavelength is roughly the size of NSs. In such case, the scalar field $\phi$ can be ignored in cosmological dynamics although one needs the cosmological constant or dark energy to explain the accelerated expansion of the universe. For this reason, a constraint on $\lambda$ is irrelevant to cosmological observations in our case.

The variation with respect to the metric gives the Einstein equation,
\ba
&&\mpl^2 \Bigl[ F(\phi) G_{\mu\nu} -
F_{,\phi} \left(\nabla_{\mu}\nabla_{\nu} \phi-g_{\mu\nu} \,\square \phi \right)
-F_{,\phi\phi} (\nabla_{\mu} \phi \nabla_{\nu}\phi - g_{\mu\nu} \nabla^{\alpha}\phi\nabla_{\alpha}\phi) 
\Bigr] \notag\\
&&~~~~~~~~~=
{1\over 2}(1-6Q^2) F(\phi)  \Bigl[
2\nabla_\mu \phi \nabla_\nu \phi
-
g_{\mu\nu} \nabla_\alpha \phi \nabla^\alpha \phi 
\Bigr]
- g_{\mu\nu} V(\phi)
+ T_{\mu\nu}\,,
\ea
where the energy-momentum tensor is defined as in usual, 
$T_{\mu\nu} = - (2 / \sqrt{-g})(\delta S_{\rm m}/ \delta g^{\mu\nu})$,
$F_{,\phi}= \partial F/ \partial \phi$, and $F_{,\phi\phi}= \partial^2 F/ \partial \phi^2$.
The Euler-Lagrange equation for the scalar field $\phi$ is given by
\ba
(1-6Q^2) F(\phi) \square \phi 
+ {1 \over 2}F_{,\phi}  \Bigl[ \mpl^2  R +(1-6Q^2) \nabla_\alpha  \phi \nabla^\alpha \phi  \Bigr] -V_{,\phi} =0 \,.
\label{EOMphiC}
\ea
The Ricci scalar in (\ref{EOMphiC}),
can be removed by taking the trace of the Einstein equation.
As one can see, with the choice of $Q$ specified by \eqref{potentialV}, the kinetic term in \eqref{EOMphiC} vanishes. However, substituting the trace of Einstein equation back into the scalar field equation, it can be easily seen that the scalar field sourced by the trace of the energy-momentum tensor of the matter field, and the kinetic term of the scalar field appears after this substitution. As in general relativity, the Bianchi identity provides the well-known form of the energy-momentum conservation, 
\ba
\nabla^\mu T_{\mu\nu} = 0 \,.
\label{EMcon}
\ea

\subsection{Modified Tolman-Oppenheimer-Volkoff (TOV) equation}

In this subsection, we summarize the modified  Tolman-Oppenheimer-Volkoff (TOV) equations and boundary conditions. 
We adopt the following static and spherically symmetric background metric, 
\ba
{\rm d}s^2=-f(r) \rd t^2+h^{-1}(r) \rd r^2 
+r^2 \left( \rd \theta^2+\sin^2 \theta\,\rd \varphi^2 \right)\,,
\label{metJ}
\ea
and consider a perfect fluid, which is given by $T^\mu_{~\nu} = {\rm diag} (-\rho(r), P(r), P(r), P(r)) $. Here, $\rho$ is the energy density and $P$ is the pressure for the matter.
The energy-momentum conservation \eqref{EMcon} gives the continuity equation
\ba
P'+\frac{f'}{2f} \left( \rho+P \right)=0\,, 
\label{continuity}
\ea
where a prime represents the derivative with respect to $r$.
After solving the Einstein equations and the scalar field equation in terms of $f'$, ${\cal M}$, and $\phi''$, we get \cite{Kase:2019dqc}
\ba
\frac{f'}{f} &=& -\frac{2M_{\rm pl}^2 (h-1)
	-2F^{-1} r^2 (P-V)+hr \phi' [(6Q^2-1)r \phi'-8Q M_{\rm pl}]
}{2h r M_{\rm pl} (M_{\rm pl}-Qr \phi')}\,,\label{eq1}\\
{\cal M}' &=& 4\pi F^{-1} r^2 \left[ (1-2Q^2) \rho+6Q^2 P
+(1-8Q^2)V-2QM_{\rm pl} V_{,\phi} \right] \nonumber \\
& & +\phi' \frac{2QM_{\rm pl}{\cal M}+8Q M_{\rm pl} \pi r^3 F^{-1}
	(P-V)+r\phi'(4 \pi r M_{\rm pl}^2-{\cal M})(1+2Q^2)}{2M_{\rm pl}(M_{\rm pl}-Qr \phi')}\,,\label{eq2}\\
\phi'' &=& -\frac{\phi'}{2M_{\rm pl}^2 rh} 
\left[ 2(h+1)M_{\rm pl}^2+r^2 F^{-1} \{ 
P-\rho+2 QM_{\rm pl} V_{,\phi}-2V
+2(\rho-3P+4V)Q^2 \} \right] \nonumber \\
& &+\frac{1}{M_{\rm pl}hF}
\left[ 4QV+V_{,\phi}M_{\rm pl}
+Q(\rho-3P) \right]\,,\label{eq3}
\ea
where we introduced the mass function ${\cal M}(r)$,
\ba
h(r) = 1- {2 G {\cal M } (r) \over r}. 
\label{defM}
\ea

\begin{figure}[t]
\begin{center}
\includegraphics[width=0.6\linewidth]{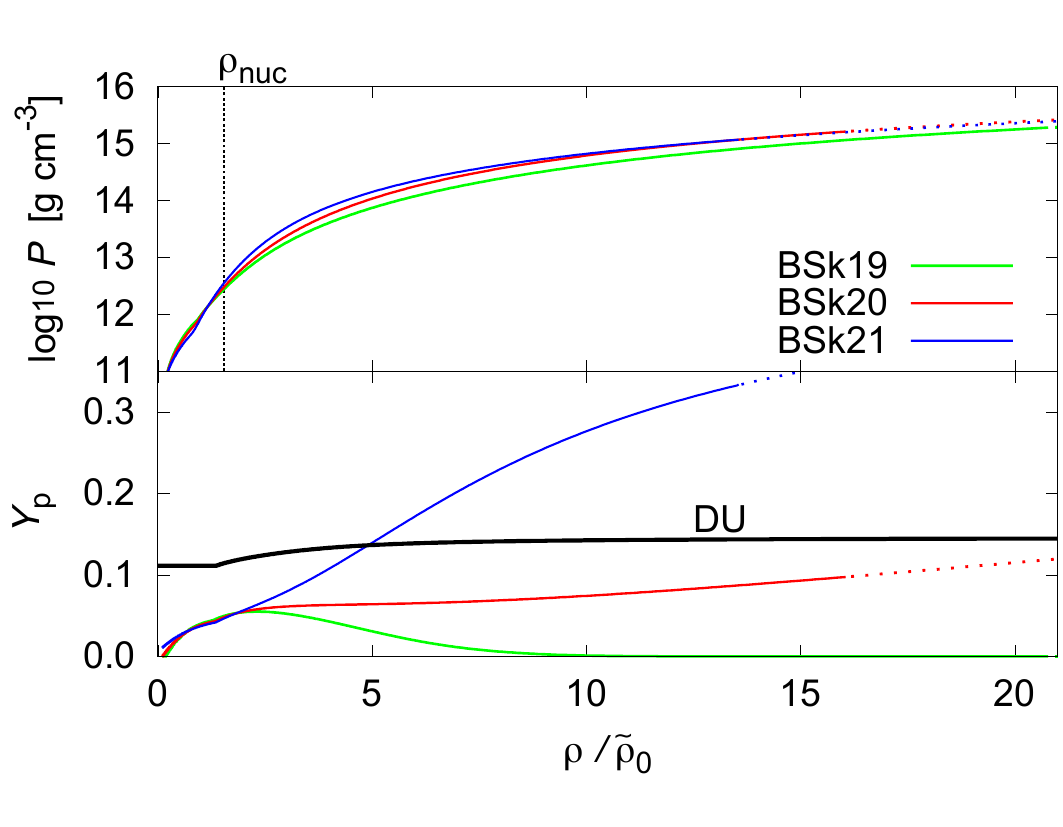}
\caption{Top: Density -- Pressure relation with the EOSs, BSk19(green), BSk20(red), and BSk21(blue). The vertical dotted line shows the proximate nuclear saturation density $\rho_{\rm nuc}$. Bottom: Proton fraction $Y_p$ as a function of density with the three EOSs. The black curve indicates the DU threshold via electrons with BSk21 EOS. All solid curves with each EOS are in stable NS region while dotted curves are in unstable region.}
		\label{fig:fig10}
	\end{center}
\end{figure}

\subsection{Equation of state}
\label{sec:EOS}

We adopt the three EOSs constructed on the Brussels--Montreal-Skyrme (BSk) functionals \cite{Skyrme1956}, which are the Hatree--Fock--Bogoliubov (HFB) mass models with `usual' Skyrme effective interaction including extra two terms of density--dependent generalizations, BSk19, BSk20 and BSk21~\cite{Goriely2010}. These functionals are fitted to the experimental nuclear masses from Atomic Mass Evaluation (AME) in 2010, and the experimental constraints of the symmetry--energy parameters, including the slope parameter $L$, from measurements of heavy--ion collisions and neutron--skin thickness.  Moreover, these EOSs are accurately fitted to realistic pure--neutron matter (PNM) EOS; BSk19, BSk20, and BSk21 EOSs are fitted to FP~\cite{Friedman1981},  APR~\cite{Akmal1998}, and LS2~\cite{Li2008} PNM EOSs, respectively, which are based on a variational method with the realistic 2--body interaction with 3--body force. Thus, the EOSs constructed with BSk functionals are successful to describe properties of both measured nuclei in the ground state and the infinite nuclear matter. To obtain the basic properties of cold NS matter, we use the public code\footnote{\url{http://www.ioffe.ru/astro/NSG/BSk/index.html}} which fully reproduces the three EOSs~\cite{Potekhin2013}. 

In Fig.~\ref{fig:fig10}, we show the EOS softness and the distribution of proton fraction $Y_p$ as functions of the density.
As one can see, the softest EOS among three EOSs is BSk19, while the stiffest EOS is BSk21. The value of $Y_p$ of the BSk21 is much higher than that of the other EOSs for $\rho\gtrsim 2\tilde{\rho_0}$. We note that the values of $Y_p$ of the BSk19 and BSk20 are clearly lower than DU threshold for all region, which
predicts non-rapid cooling of NS. These profiles of pressure and $Y_p$ distribution can be explained by the difference of the slope parameter of the symmetric energy defined by $L = \rho_{\rm nuc}\left.\frac{\partial S(\rho)}{\partial \rho}\right|_{\rho=\rho_{\rm nuc}}$, where $S(\rho)$ is the symmetry energy and $\rho_{\rm nuc}$ is nuclear saturation density. The value of $L$ calculated in Ref.~\cite{Goriely2010} is $L = 31.9~{\rm MeV}$ for the BSk19, $L = 37.4~{\rm MeV}$ for the BSk20, and $L = 46.6~{\rm MeV}$ for the BSk21,
whose values are consistent with the behaviors of the pressure and $Y_p$ distributions in Fig.~\ref{fig:fig10}.
Thus $L$ is an indicator of the relevancy for the EOS softness and the threshold mass of the nucleon DU process (hereafter $M_{\rm DU}$). For more precise predictions, it is necessary to know the higher-derivative terms of the symmetry energy with respect to the finite density and $Y_p$, where the interactions between nucleons are still unclear. Therefore, the EOS itself should have some uncertainties of the order of $10\%$ in $\rho$--$P$ relation. 
As we will see in the next section, the fifth force is sensitive to $\rho - 3P$, and this will therefore change the mass-radius relations. However, our method in the present paper should be still valid for other EOSs.

\subsection{Numerical solutions}

\begin{figure}[ht]
\begin{tabular}{cc}
\begin{minipage}{0.49\linewidth}
\begin{center}
\includegraphics[width=1.\linewidth]{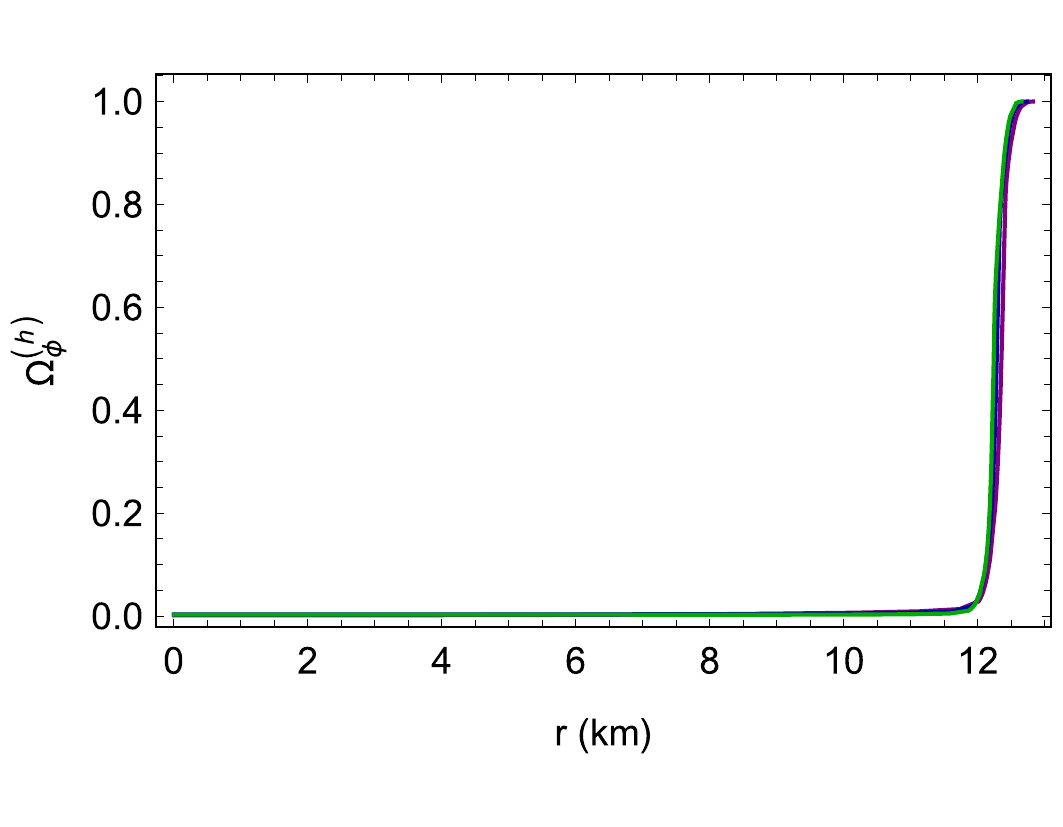}
\end{center}
\end{minipage}
\begin{minipage}{0.49\linewidth}
\begin{center}
\includegraphics[width=1.\linewidth]{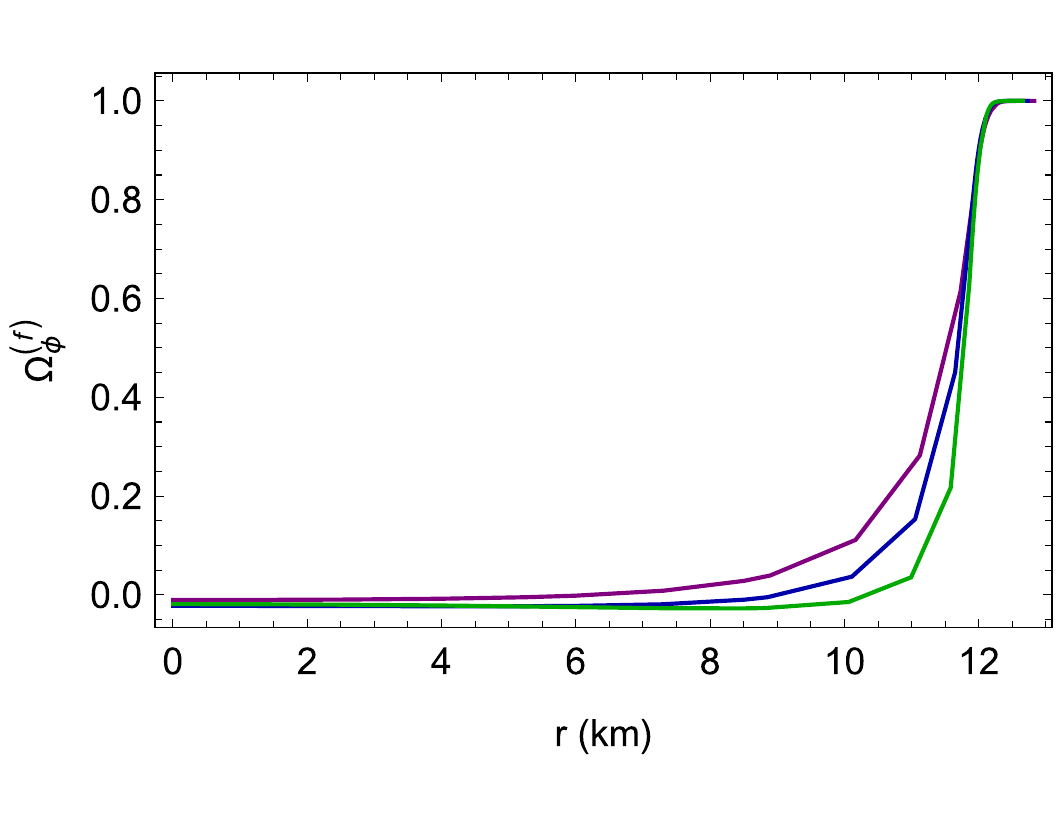}
\end{center}
\end{minipage}\\
\vspace{5mm}
\\ 
\begin{minipage}{1.\linewidth}
\begin{center}
  \includegraphics[width=0.5\linewidth]{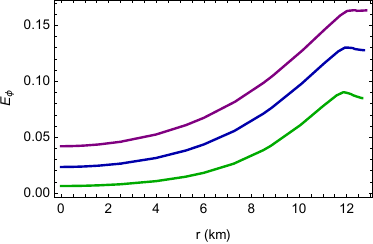}
\end{center}
\end{minipage}
\end{tabular}
\caption{$\Omega_\phi^{(f)}$ (upper left panel), $\Omega_\phi^{(h)}$ (upper right panel), and $E_\phi$ (lower panel)
		as a function of radius $r$ in the unit of {\rm km}. The purple, blue, and green line correspond to ${\tilde \lambda}=10^3, 10^4, 10^5$, respectively. The central density is given by $\rho_0 = 8.1 \times 10^{14} {\rm g/cm^3}$ and the EOS is adopted BSk21.
}
\label{fig:chame}
\end{figure}

Here, we give an overview of the structure of NSs by solving the modified TOV equations with the EOSs, BSk19, BSk20, and BSk21. In order to numerically solve these equations, 
we impose the following boundary conditions such that the regularities at the center of NSs are satisfied,
\ba
f' (r=0)=0\,,\qquad h'(r=0)=0\,,\qquad \phi'(r=0)=0\,, 
\qquad \rho'( r=0)=0\,.
\label{BC0}
\ea
For simplicity, we assume the asymptotic flatness at spatial infinity,\footnote{Strictly speaking, this assumption is valid only when the the cosmological dynamics of the scalar field can be ignored. We only focus on extremely large mass cases, i.e., the Compton wavelength of the scalar field is around $\lambda_{\rm C}\sim{\cal O}(1) -{\cal O}(10^3) ~{\rm km}$ (See Appendix \ref{app:chameleon} for the definition of the Compton wavelength.). This is much smaller than the present Hubble scale, and the asymptotic value of $\phi$ can be therefore approximated to be zero. On the other hand, in the massless Brans-Dicke theory studied in Appendix~\ref{app:BD}, the cosmological solution of $\phi$ has a growing mode, thus this assumption is in general invalid.} 
\ba
f(r \to \infty)=1\,,\qquad h(r \to \infty)=1\,,\qquad
\phi'(r \to \infty)=0\,,\qquad 
\phi (r \to \infty)=0\,.
\label{BC1}
\ea
Hereafter, we use the rescaled dimensionless parameter, 
\ba
{\tilde \lambda} \equiv \lambda (r_0 \mpl)^2\,, 
\label{BC2}
\ea
with $r_0 \mpl = 1.107 \times 10^{39}$. 
By imposing the boundary conditions (\ref{BC0}), 
we numerically solve Eqs.~(\ref{continuity})-(\ref{eq3}) with the EOS 
outward from the center of star ($r=0$) in order to clarify the profiles of 
$f, {\cal M}, \phi, \rho,$ and $P$ inside the star. 
The star radius $r_s$ is determined by the condition $P(r_s)=0$, and we identify 
the mass of star $M_s={\cal M}(r_s)$. Outside the star, 
we simply set $\rho=P=0$ and solve Eqs.~(\ref{eq1})-(\ref{eq3}) for $f, {\cal M}, \phi$ 
up to the sufficient large distance, e.g., $r=10^5r_s$, at which both $\phi$ and $\phi'$ 
sufficiently approach 0. Here we note that the asymptotic values of $f$ and $\phi$ at 
$r\to\infty$ do not satisfy the other boundary conditions (\ref{BC1}) in general. 
However, we can shift the asymptotic value of $f$ to 1 by virtue of time reparametrization 
invariance. Regarding the scalar field, we resort to a shooting method in order to find 
the initial value $\phi(r=0)$ which eventually satisfies the boundary condition $\phi(r\to\infty)=0$. 
See \cite{Kase:2019dqc} for more details of numerical procedure. 

In this work, following the Ref.~\cite{Kase:2019dqc}, we fix the dimensionless parameter $\tilde \lambda$ to be $0$, $10^3$, $10^4$, $10^5$. By choosing these values, the corresponding Compton wavelength of the scalar field can be comparable with the size of NSs, implying that the fifth force effect can significantly affect the structure of NSs. However, these parameters are unfortunately inconsistent with the constraint by the E\"{o}t-Wash torsion balance experiment~\cite{Upadhye2012}, $\lambda\gtrsim\mathcal{O}(1)$ (e.g., \cite{Burrage2016}), while it is still consistent with other astrophysical and laboratory tests \cite{Jana2019}. Nonetheless, this choice of parameters enables us to capture the effect of the fifth force in the cooling curve even for other modified gravity theories possessing the Chameleon mechanism as we will discuss in Sec. IV and V. 

Let us first discuss the behavior of the scalar field. Since the scalar field equation, \eqref{EOMphiC} or \eqref{eq3}, is complicated due to the non-minimal coupling to gravity, we work in the Einstein frame for convenience. By using the conformal transformation of the metric $(g_{\mu \nu})_{\rE}=F(\phi) g_{\mu \nu}$, one can remove the non-minimal coupling of the scalar field to gravity, and the resultant theory consists of the Einstein-Hilbert term, the canonical scalar field, and the matter field coupled to the scalar field. In Appendix~\ref{app:EF}, we have summarized  
the equation of motions for gravity, the scalar field, and the matter component, and the relation between the Jordan frame and the Einstein frame. We note that all the boundary conditions \eqref{BC0} and \eqref{BC1} and the central density $\rho_0$ are given in the Jordan frame  in numerical computation, although we discuss the behavior of the scalar field in the Einstein frame in the below.
\begin{figure}[t]
	\begin{center}
          \includegraphics[scale=0.25]{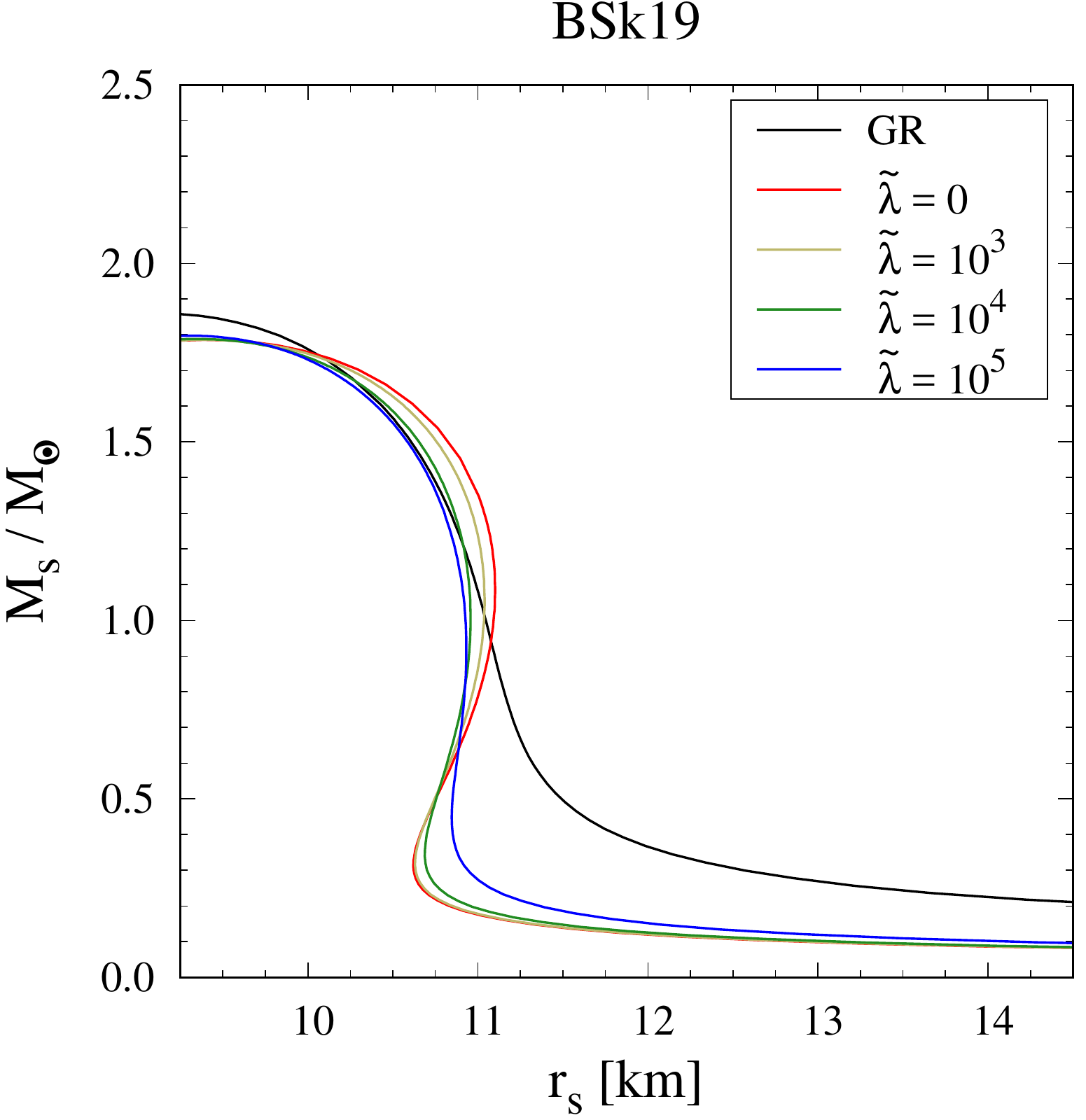}
          \includegraphics[scale=0.25]{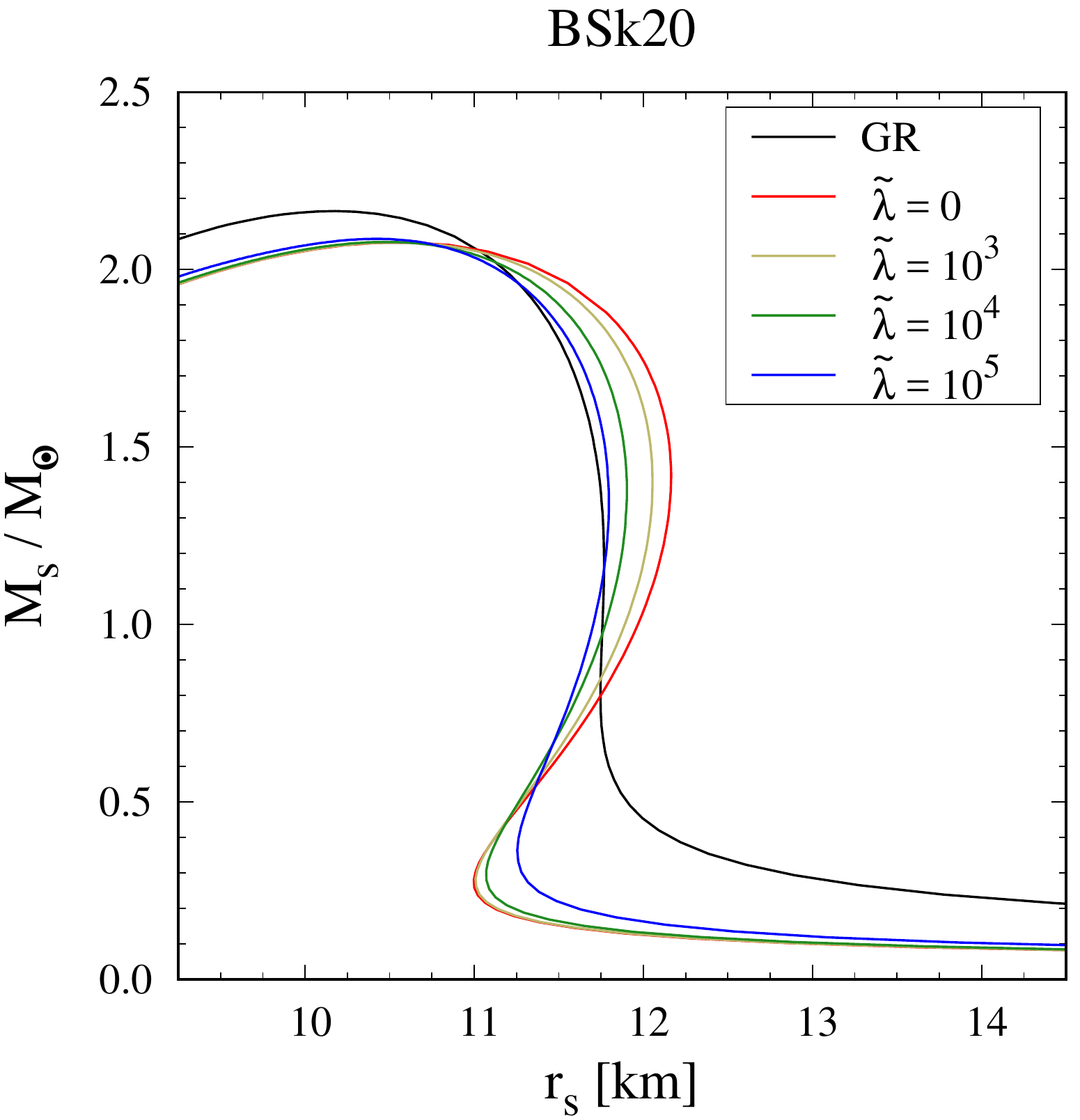}\\
           \includegraphics[scale=0.25]{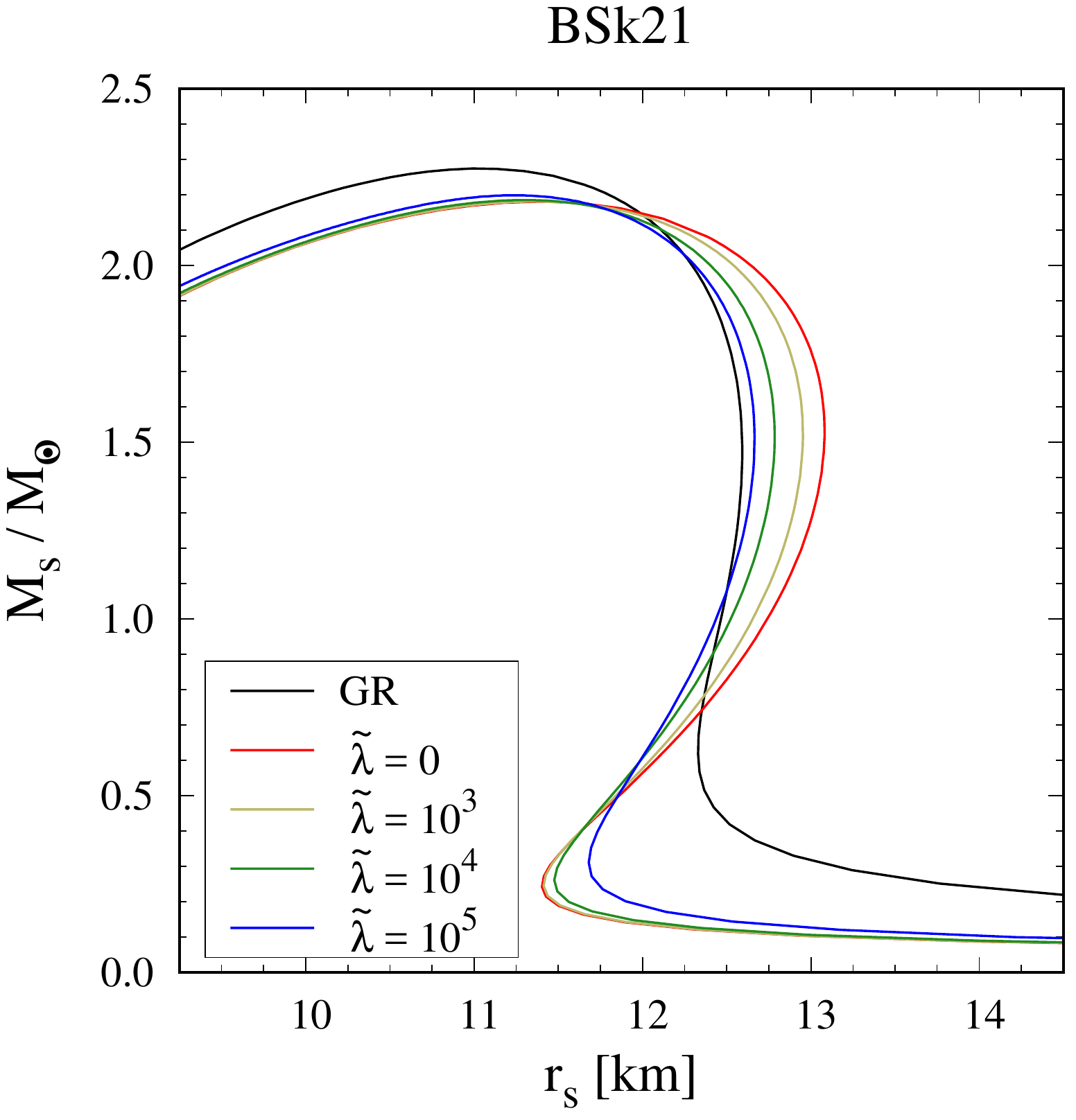}\\
	\end{center}
	\caption{\label{MR}
		Mass-radius relation for BSk19 (upper left panel), BSk20 (upper right panel), and BSk21 (lower panel) EOSs.
	}
\end{figure}

\begin{figure}[t]
	\begin{center}
               \includegraphics[scale=0.25]{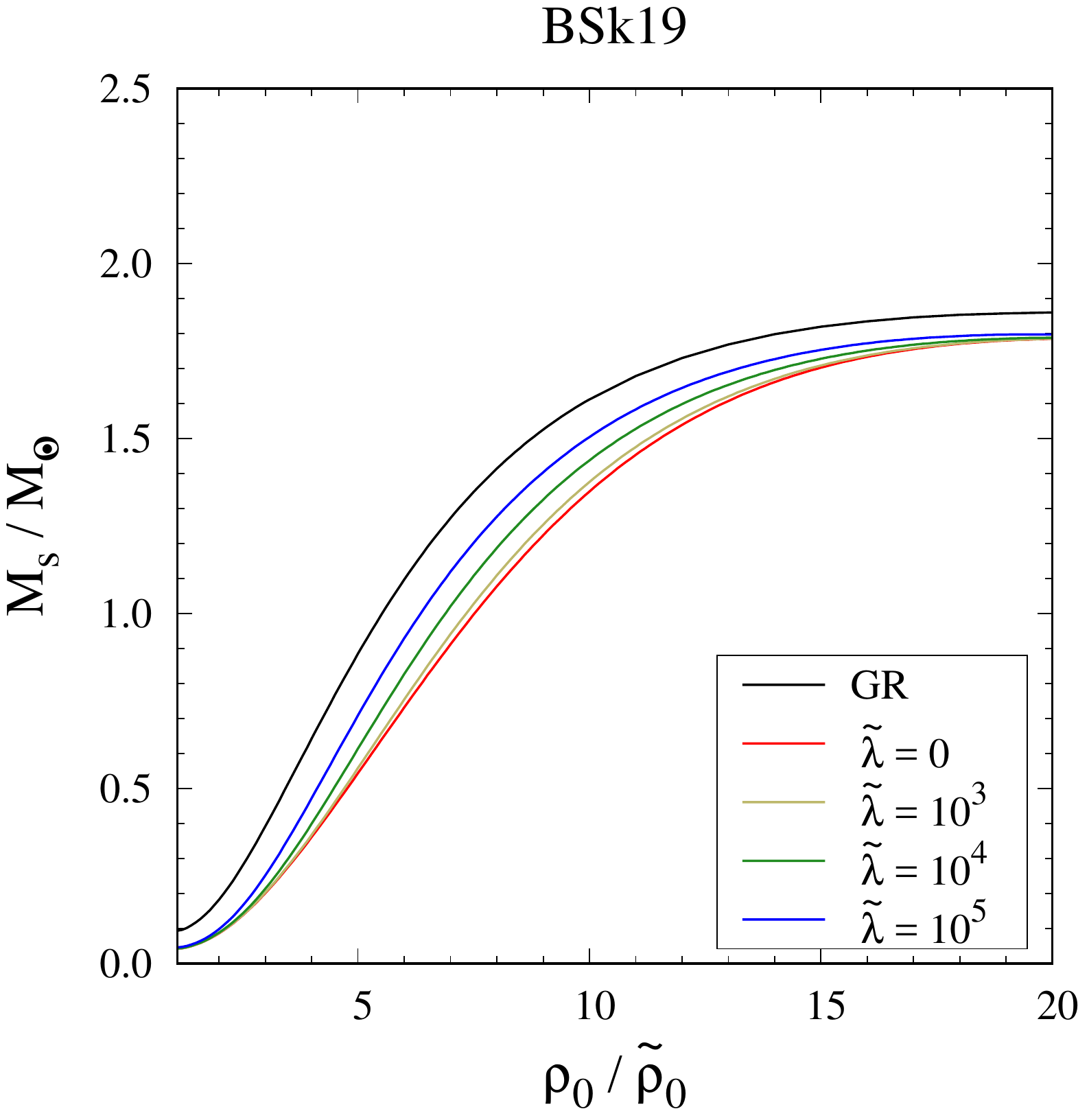}
                \includegraphics[scale=0.25]{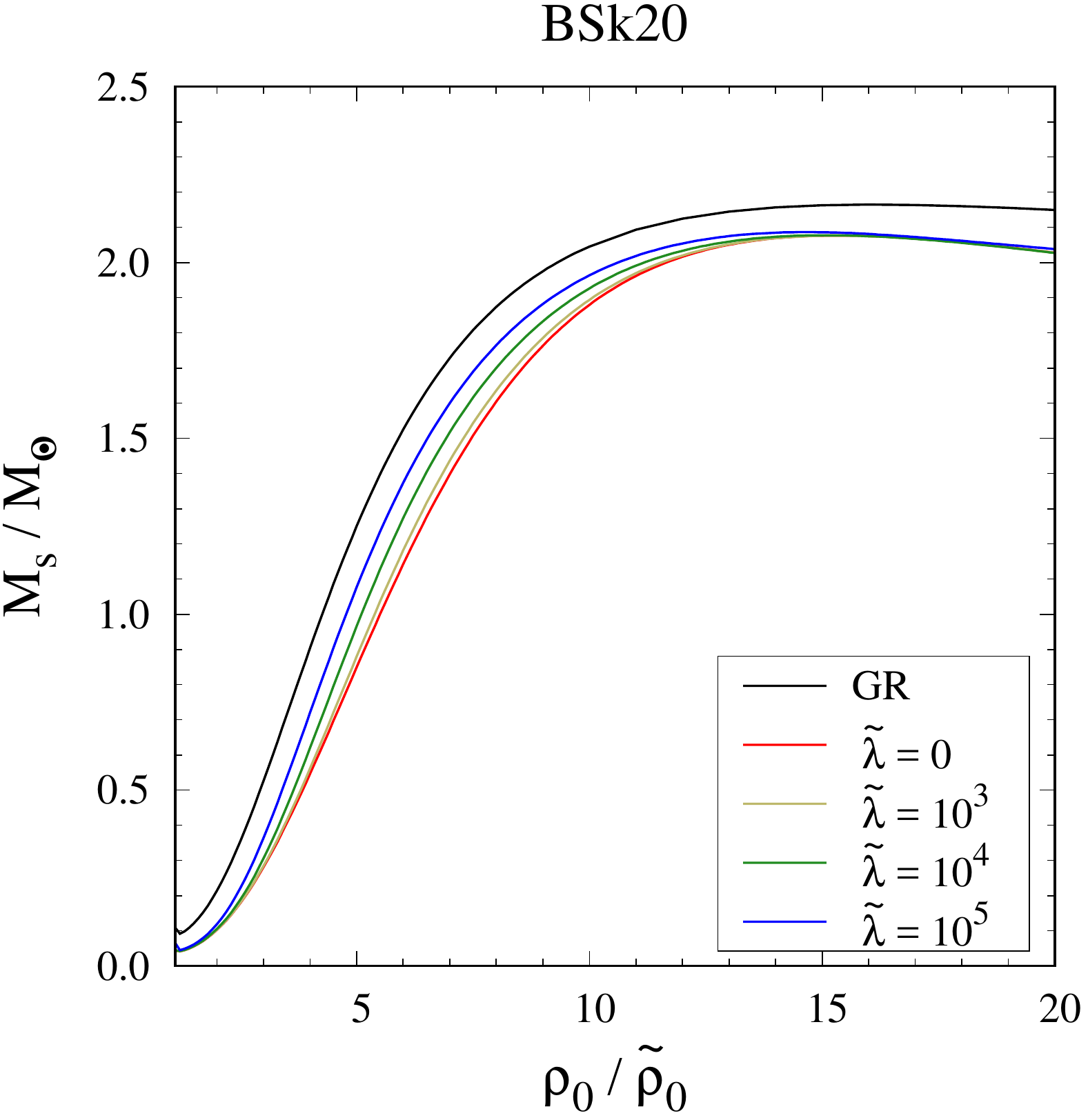}\\
                \includegraphics[scale=0.25]{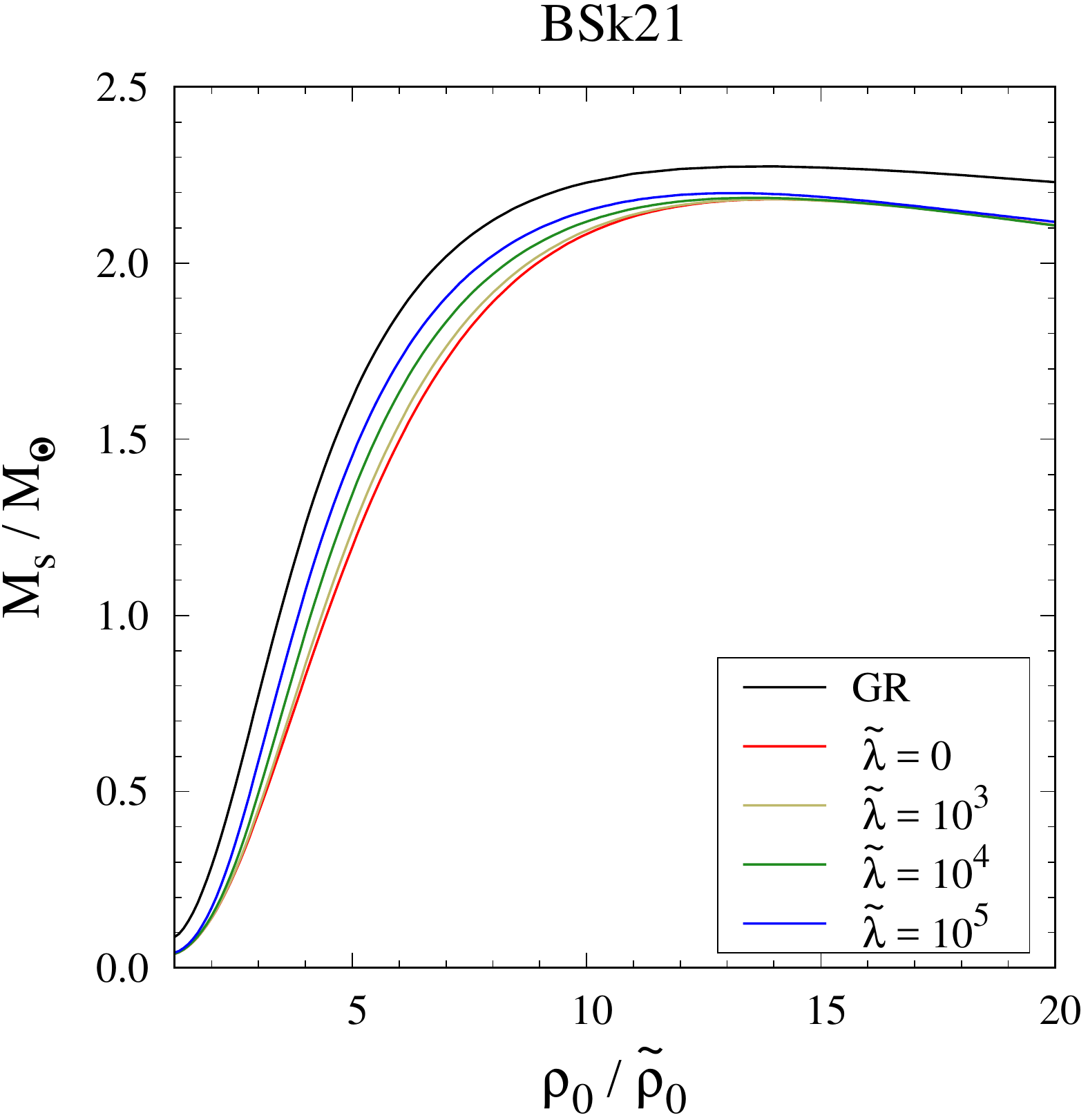}
	\end{center}
	\caption{\label{Mrho}
		Mass-central density relation for BSk19 (upper left panel), BSk20 (upper right panel), and BSk21 (lower panel) EOSs.
	}
\end{figure}

As summarized in Appendix~\ref{app:EF}, 
the scalar field obeys the Klein-Gordon equation with the effective potential given by \eqref{Veff}. Outside NSs, the effective potential is nothing but $V(\phi)$. As $\rho_\rE -3P_\rE$ increase, the minimum of the effective potential is no longer located at $\phi=0$, and the effective mass $m^2_{\rm eff} = {\rm d}^2 V_{\rm eff} / {\rm d}\phi^2 $, being dependent on $\phi$, becomes large inside NSs. When the scalar field acquires sufficiently large mass in such situation, the fifth force due to the scalar field is in general screened, and deviation from general relativity becomes relatively small.
Therefore, we naively expect that a screening of the fifth force could be effective inside NSs. 
To see the fifth force effect we introduce the following parameters in the Einstein frame as defined in 
\eqref{para1} and \eqref{para2}
\ba
\Omega_\phi^{(h)} = \frac{(T^{0}_{~0}{}^{(\phi)})_{\rm E} }{(G^0_{~0})_{\rm E} }, \qquad 
\Omega_\phi^{(f)} = \frac{(T^{1}_{~1}{}^{(\phi)})_{\rm E} }{(G^1_{~1})_{\rm E} }\,, \qquad 
E_\phi = \Biggl|
\left(\frac{\rd P_{\rE}}{\rd r_{\rE}}\right)^{-1}
\frac{Q}{M_{\rm pl}} \left( \rho_{\rE}-3P_{\rE} 
\right) \frac{\rd \phi}{\rd r_{\rE}}\Biggr|\,,
\notag
\ea
where $r_\rE$, $(T^{\mu}_{~\nu}{}^{(\phi)})_{\rm E}$, $(G^\mu_{~\nu})_{\rm E}$, $\rho_\rE$, and $P_\rE$
are the radius, the energy-momentum tensor of the scalar field, the Einstein tensor, the energy density, and the pressure in the Einstein frame. $\Omega^{(h)}_\phi$ and $\Omega^{(f)}_\phi$ characterize how much energy density or pressure of the scalar field contribute in the Einstein equations, and the  parameter $E_\phi$ characterizes the fifth force due to the scalar field, normalized by pressure-gradient, i.e., the first term in the continuity equation \eqref{continuityE}.
Namely, $E_\phi$ characterizes the ratio of the fifth force to the total force.
Roughly, we may regard the effective gravitational constant as $G_{\rm eff}\simeq (1+E_\phi)G$.
In Fig.~\ref{fig:chame}, we plot these dimensionless parameters as a function of the radius in the Jordan frame, $r$. 
For any $\tilde \lambda$, the parameter $\Omega^{(h)}_\phi$ is approximately zero well inside NSs and rapidly increases a little inside the surface of NS. On the other hand, the plot of  $\Omega^{(f)}_\phi$ shows that it can be large, for instance $\Omega^{(f)}_\phi \simeq 0.2$ at $r=10 ~{\rm km} $ for ${\tilde \lambda} = 10^3$. In addition, the effect of the fifth force is still active even well inside NSs for ${\tilde \lambda} = 10^3$,  as one can see the plot of $E_\phi$. On the other hand, for ${\tilde \lambda} = 10^5$, at sufficiently small $r$, the fifth force is negligible, thus this implies that Chameleon mechanism effectively works for sufficiently large $\tilde \lambda$ well inside NSs
\footnote{	In fact, the Compton wavelength of the effective mass for ${\tilde \lambda}=10^3$ outside and inside NSs are respectively roughly $20$ km and $8$ km. Therefore, this behavior is due to the Chameleon mechanism, not due to the large $\tilde \lambda$.}.
However, since the energy density and pressure for the matter is zero for $r \geq r_s$,  the effective mass $m_\phi$ tends to increase for $r < r_s$. This implies that the effect of the fifth force is still significant near the surface even for large $\tilde \lambda$, which can be also seen in the Fig.~\ref{fig:chame}, and this fact affects the structure of NSs. 

These arguments in the Einstein frame agree with mass-radius relation in our numerical calculation.
Fig.~\ref{MR} shows the mass $M_s$ (normalized by the solar mass $M_{\odot} = 1.9884 \times 10^{33}$ g) as functions of the radius $r_s$ for  BSk19 (upper left panel), BSk20 (upper right panel),
and BSk21 (lower panel) EOSs. 
Even for ${\tilde \lambda} = 10^5$, the fifth force effect near the surface of NS significantly affects MR relation. 
As one can see, the maximum mass of NSs decreases as the parameter $\tilde \lambda$ decreases. Since the fifth force due to the scalar field is attractive and the net gravitational force from the metric  $g_{\mu\nu}$ and the scalar field $\phi$ also becomes larger. Based on this fact, NSs tend to be compact for small $\tilde \lambda$, which can be also seen from the Fig.~\ref{Mrho}, and this is why the maximum mass of general relativity is the largest value. 
This feature is similar to the gas density profile of galaxy clusters in an $f(R)$ gravity model, as found in
Refs.\cite{terukina1,terukina2}.

\section{SETUP OF NEUTRON STAR COOLING}
\label{coolingsec}

INS with the low magnetic field has no heating source, and after being born by a supernova explosion, INS cools down by the emission process of neutrino at the early stage ($t\lesssim 10^{4-5}~{\rm yr}$) and by that of photons at the late stage ($t\gtrsim 10^{5}~{\rm yr}$). The emissivities of neutrinos and photons are therefore important to describe the thermal evolution of INS. In this section, we give a brief review of physics about the neutrino and photon energy loss.

The most important factor for comparing with the observations of INS is the neutrino emission process, which has been studied by many works (for review, see Ref. \cite{Yakovlev2001}).
We solve the thermal evolution of INS in the Jordan frame, then we assume that the radiative processes through
photons and neutrinos are not affected by the modification of the gravity.
In general, the neutrino emission processes are classified into fast and slow cooling processes. Within the standard matter, $n,~p,~e,~\mu$, the fast cooling process is the Direct Urca (DU) process, which is the forward and inverse $\beta$ decay. The neutrino emissivity of the DU process is given by \cite{Coleman1979,Lattimer1991a}
\be
\epsilon^{\rm Fast}_{\nu} = 4.00\times 10^{27} \left(\frac{m_n^*}{m_n}\right)\left(\frac{m_p^*}{m_p}\right) \left(\frac{\rho_{\rm B}}{\rho_{\rm nuc}} \right)^{2/3} T_9^6 \left (\sum_{l = e,\mu} Y_l^{1/3}\Theta_{npl} \right)~~{\rm erg~cm^{-3}~s^{-1}}\,, \label{eq:4a.1}
\ee
where $m_n^*/m_n$ and $m_p^*/m_p$ are effective mass ratio of neutrons and protons, respectively, which depend on the EOS. Here $\rho_{\rm B}$ is the baryon density, $T_9$ is the temperature in units of $10^9~{\rm K}$, and $Y_l$ is the lepton fraction, $\Theta_{npl}$ denotes the step function:
$\Theta_{npl} = 1$ when the momentum conversation is satisfied $\bm{k}_{{\rm F}n} = \bm{k}_{{\rm F}p} + \bm{k}_{{\rm F}l}$, while $\Theta_{npl} = 0$ when $\bm{k}_{{\rm F}n} \neq \bm{k}_{{\rm F}p} + \bm{k}_{{\rm F}l}$,
where $\bm{k}_{{\rm F}n}$, $\bm{k}_{{\rm F}p}$, and $\bm{k}_{{\rm F}l}$ are the wavenumber of neutrons, protons, and leptons, respectively. From the energy-momentum conservation of $\bm{k}_{{\rm F}n}$, $\bm{k}_{{\rm F}p}$, and $\bm{k}_{{\rm F}e}$, the proton fraction $Y_p$ for the DU process being effective satisfies the following condition:
\be
Y_p\ge Y^{\rm DU}_p = \frac{1}{1+\left(1 + x_e^{1/3}\right)^3}~, \label{eq:4a.2}
\ee
where $Y^{\rm DU}_p$ is the net threshold $Y_p$ of DU and $x_e = Y_e/\left(Y_e + Y_{\mu} \right)$. In the very high-density regions, (\ref{eq:4a.2}) becomes $Y^{\rm DU}_p \gtrsim 0.1477$ because of the condition $Y_e \simeq Y_{\mu}$. Whether the DU process occurs or not is determined by the proton fraction in the central region of INS, which may be changed by 
gravity and EOS models. 
Thus, the gravity and the EOS models are
significant for the cooling of INS. 

Generally, the DU process is prohibited for INS with light masses, in which
the thermal evolution of INS is determined by the slow cooling processes, the Modified Urca (MU) process and the neutrino bremsstrahlung radiation in nucleon-nucleon collisions \cite{Friman1979,Yakovlev1995}. 
The MU process is the reaction of $n+N\rightarrow p + N + l+ \bar\nu_l,~p+N+l\rightarrow n+N+\nu_l$, where $N$ is neutron or proton, although the DU process is $n\rightarrow p+l+\bar \nu_l$, ~$p+l\rightarrow n+\nu_l$.
The bremsstrahlung is the neutrino radiation emitted when electrons or muons are braked by the Coulomb field of protons. These emissivities can be approximately expressed as
\be
\epsilon^{\rm Slow}_{\nu} \approx 10^{19-21}\left(\frac{\rho_{\rm B}}{\rho_{\rm nuc}}\right)^{1/3} T_9^8 ~~{\rm erg~cm^{-3}~s^{-1}}\,. \label{eq:4a.3}
\ee
Here, we ignore the dependences of effective mass ratio and fraction of particles in (\ref{eq:4a.3}) though we consider them in numerical calculation. Compared to (\ref{eq:4a.1}), we see that the slow cooling processes are much weaker than the DU process, but most of the reactions always occur above the threshold density of neutron drip\footnote{The MU process with the proton branch does not occur for $Y_p\le 1/65$.}. Therefore, the slow cooling processes, especially the modified Urca process which is stronger than the bremsstrahlung \cite{Yakovlev1995}, are important for describing thermal evolution of all INSs. 


\begin{figure}[t]
    \centering
    \includegraphics{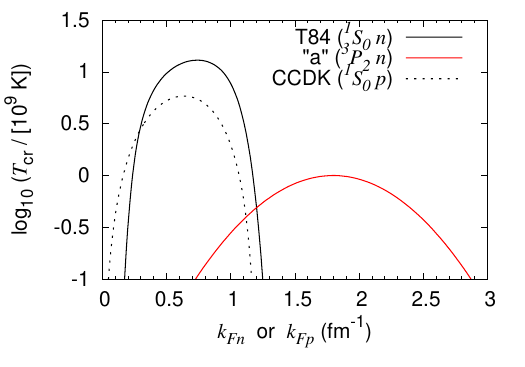}
    \caption{Adopted nucleon superfluid models, that is, $T_{\rm cr}$ dependence on the Fermi wave number of neutrons or protons. Solid curves denote the neutrons superfluidity while dotted protons superconductivity. The adopted superfluid models are as follows: Ref.~\cite{Takatsuka1972} for the ${}^1S_0$ neutrons superfluidity, Ref.~\cite{Chen1993} for the $^1S_0$ protons superconductivity, and Ref.~\cite{Page2004} with the ``a" model for ${}^3P_2$ neutrons superfluidity.}
    \label{fig:sfmodel}
\end{figure}

In modern cooling theory, however, the above cooling processes are not sufficient because neutrons or protons are highly expected to be in a superfluid state due to low-temperature environment. Specifically, the ${}^1S_0$ neutrons superfluidity is contained in the inner crust, and the ${}^3P_2$ neutrons superfluidity and ${}^1S_0$ protons superconductivity exist in the core.
According to the BCS theory~\cite{BCS1957}, such nucleons make their pairs which can be expressed as following reaction:
\begin{align}
N+N\rightarrow [NN] + \nu + \bar{\nu}. \nonumber
\end{align}
Here $N$ is a neutron or proton. $[NN]$ denotes the $N-N$ Cooper pair and behaves as a bosonic particle. Each nucleon superfluidity has their superfluid transition temperature $T_{\rm cr}$, but $T_{\rm cr}$ dependence of the density still has large uncertainties above all for the ${}^3P_2$ neutrons superfluidity. In this work, we adopt the superfluid models as shown in Fig.~\ref{fig:sfmodel}; T84 model for neutrons ${}^1S_0$~\cite{Takatsuka1972}, `a" model for neutrons ${}^3P_2$~\cite{Page2004}, and CCDK model for protons ${}^1S_0$~\cite{Chen1993}.

The important superfluid effect on NS cooling curve is the pair breaking formation (PBF) process. By making or breaking Cooper pairs, additional neutrino emissions occur with above all $T\approx 0.5T_{\rm cr}$, and the emissivities are roughly expressed as follows~\cite{Yakovlev1995}: 
\be
\epsilon^{\rm PBF}_{\nu} \approx 10^{21-22}\left(\frac{\rho_{\rm B}}{\rho_{\rm nuc}}\right)^{1/3} T_9^7\tilde{F}_i\left(\frac{T}{T_{\rm cr}}\right) ~~{\rm erg~cm^{-3}~s^{-1}}\,, \label{eq:4a.3+}
\ee
where $\tilde{F}_i$ is the control function with $T/T_{\rm cr}$, which depends on whether the state of nucleons is singlet ($i=s$) or triplet ($i=t$), and $\tilde{F}_s$ and $\tilde{F}_t$
are given in Ref.~\cite{Yakovlev1999}. Compared with the emissivity of slow cooling processes formulated as Eq.~(\ref{eq:4a.3}), the PBF process is stronger by 1$\sim$2 order of magnitude, depending on the temperature. Without any fast cooling processes, therefore, the PBF process is dominant for INS cooling curves. Thus, nucleon superfluidity effect is though to be an essential factor to consider cooling curves in modern cooling theory. 


The photon cooling is dominant for the thermal evolution of INS at the late stage.
Assuming the blackbody emission, the photon luminosity is given by 
\be
L_{\gamma}(r=r_s) = 7\times 10^{36}~{\rm erg~s^{-1}}\left(\frac{r_s}{10~{\rm km}}\right)^2aT^4_{{\rm eff},7},
\label{eq:4a.4}
\ee
where $a$ is the {\it Stefan-Boltzmann} constant, and $T_{{\rm eff},7}$ is the effective temperature in units of $10^7~{\rm K}$. $L_{\gamma}$ depends on the surface compositions of the envelope of INS. The surface
compositions reflect the relation between surface temperature and interior temperature at $\rho = 10^{10}~{\rm g~cm^{-3}}$. 
The reason why the surface compositions affect the cooling curves is ascribed to the opacities,
which include the radiative opacity and conductive opacity  mainly due to electrons. We simply adopt
the GPE relation \cite{Gudmundsson1983}, which assumes the envelope only with ${}^{56}{\rm Fe}$. 

We have assumed that the emissivities by photons and neutrinos as well as the transport equation are not affected by the modification of the gravity in the Jordan frame because the equation of motion of the matter is not directly coupled to the scalar field, which is clearly seen in Eq. (\ref{action}) or Eq. (\ref{actfR}). Then, we use the following cooling transport equations for thermal evolution of NS without heating source \cite{Thorne}
\be
C_V\frac{\partial T}{\partial t} = -\frac{1}{f^{1/2}}\left \{ \frac{h^{1/2}}{4\pi r\rho}\frac{\partial\left(L_{\gamma}f\right)}{\partial r} + f\epsilon_{\nu} \right \}\label{eq:4a.6},
\ee
where $C_V$ is the specific heat, and $\epsilon_{\nu}$ is the total neutrino emission including \equref{4a.1} and \equref{4a.3}. The boundary conditions for $T$ and $L_{\gamma}$ are given by $L_{\gamma}(r=0) = 0$ from  \equref{4a.4} and $T(t=0) = 10^{10}~{\rm K}$, but the choice of the initial temperature with $T(t=0)\gtrsim 10^{8.5}~{\rm K}$ does not affect cooling curves with $t\gtrsim 10^{1-2}~{\rm yr}$ at all.  After giving the initial temperature, we calculate the cooling curves by the low surface temperature of $T_{\rm eff}^{\infty}=10^5~{\rm K}$.

In our numerical computation to find the solution to \equref{4a.6}, we use the Henyey scheme~\cite{Henyey1964}
after solving the NS structure from Eqs.~(\ref{continuity})-(\ref{eq3}).
Then, we obtain theoretical cooling curves which describes the thermal evolution of INS.
For numerical calculation of INS cooling, we use the public code \verb|NSCool|\footnote{\url{http://www.astroscu.unam.mx/neutrones/NSCool/}}~\cite{Page2016conf} which has already included the neutrino and photon emissions.

\begin{figure}[t]
\begin{center}
\includegraphics[width=0.7\linewidth]{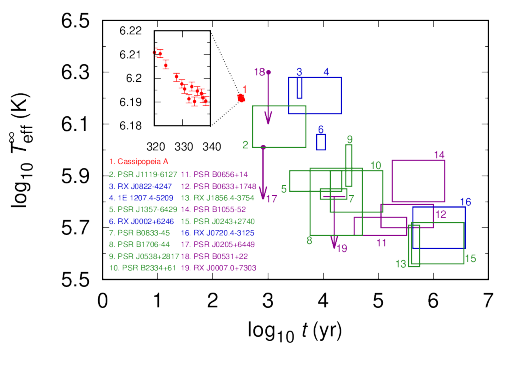}
\caption{Age and effective temperature for a distant observer of INSs adopted from Ref. \cite{Lim2017}. The difference of the color indicates that of models of the atmosphere; carbon atmosphere for the star with red color, hydrogen atmosphere with blue, magnetized hydrogen atmosphere with green, and black-body models with purple. The insets in this upper-left corners indicate magnification of the data with Cassiopeia A and the continuous data are plotted in linear axis about the age.}
\label{fig:fig11}
\end{center}
\end{figure}

To examine the validity of the models for the thermal evolution of INS, it is useful to compare the theoretical cooling curves with observational data of the age and the surface temperature. The age is mainly estimated by two method: One is based on a measurement of spin-down rate of a pulsar due to the braking by a rotating magnetic filed. The other is based on estimating the ``kinetic age'' by connecting the transverse velocity to a distance from the geometric central position of a supernova remnant. Generally, though the ``kinetic ages'' are not always known, it is thought to be more reliable than the method of measuring the spin-down rate. The surface temperature is estimated by measurements of X-ray flux. For the estimation, the uncertainty comes from modeling of the atmosphere of NS, which significantly affects the net X-ray flux. We adopt the data of Ref. \cite{Wijngaarden2019} for Cassiopeia A and other 18 INS in Ref. \cite{Lim2017}. The data of 19 INS are plotted in Fig.~\ref{fig:fig11}. 
Cassiopeia A, a young supernova remnant with the age around $t\approx 340~{\rm yr}$, is especially important because the uncertainties of the age and the surface temperature are much smaller than those of the other observations. The surface temperature of Cassiopeia A has been measured for about 20 years by Chandra X-ray detectors. However, we should note that the results may contain systematic errors because the observed decay rate
depends on the Chandra X-ray detectors and modes~\cite{Elshamouty2013,Posselt2018}. 
Constraint from the observational data for Cassiopeia A has been investigated in Refs.~\cite{Page2011,Shternin2011, Noda2013, Hamaguchi2018}.

\section{Cooling Curves without Nucleon Superfluidity}
\label{cool}

\begin{figure}[htbp]
\begin{center}
\includegraphics[width=0.9\linewidth]{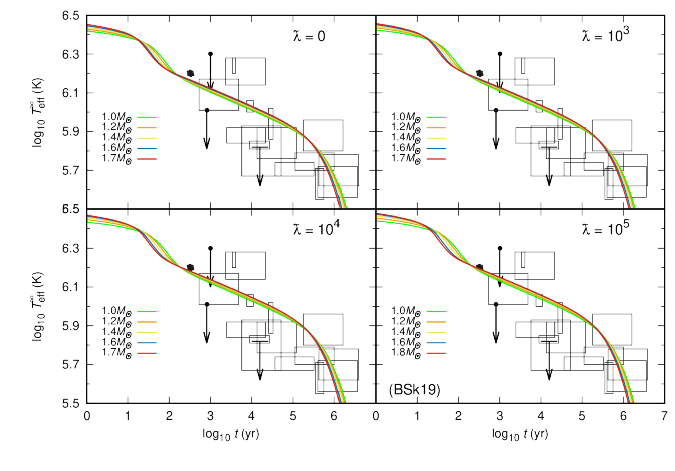}
\caption{Cooling curves using the BSk19 EOS. }
\label{fig:fig15}
\vspace{1cm}
\includegraphics[width=0.9\linewidth]{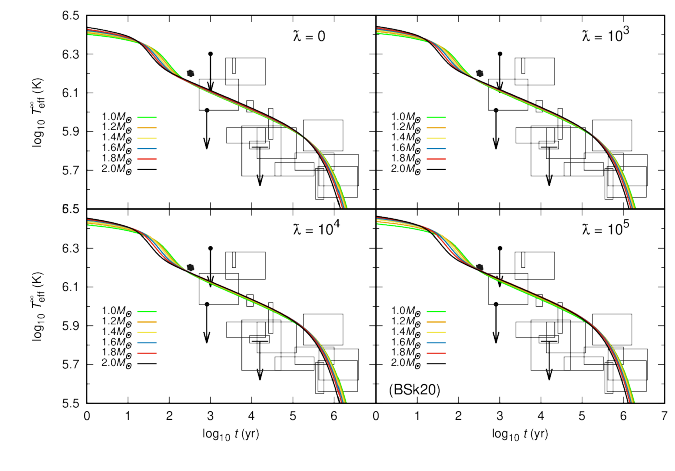}
\caption{Same as Fig.~\ref{fig:fig15}, but with the BSk20 EOS}
\label{fig:fig16}
\end{center}
\end{figure}

\begin{figure}[htbp]
\begin{center}
  \includegraphics[width=0.9\linewidth]{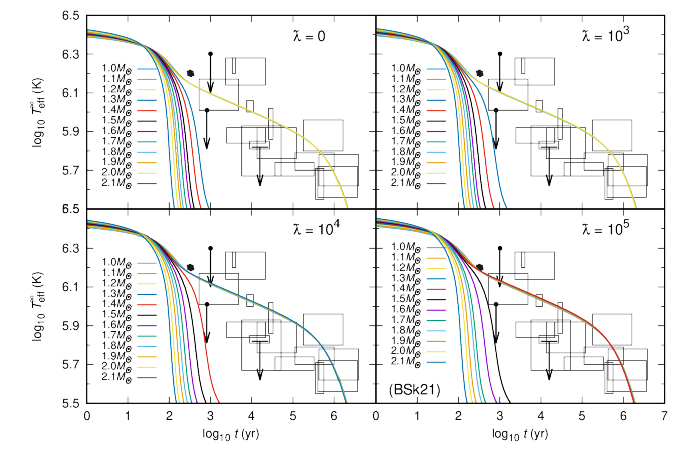}
\caption{Same as Fig.~\ref{fig:fig15}, but with the BSk21 EOS. }
\label{fig:fig17}
\end{center}
\end{figure}

\begin{figure}[htbp]
\begin{center}
\includegraphics[width=1\linewidth]{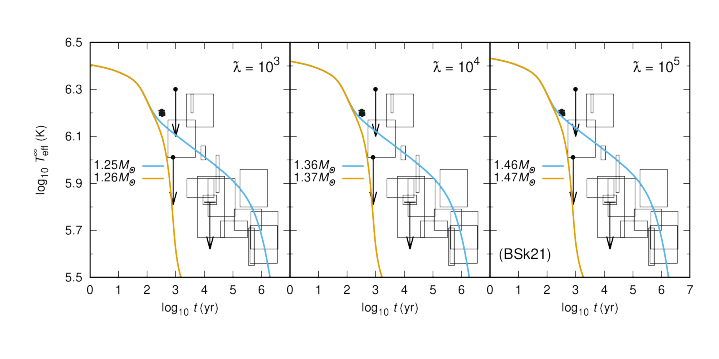}
\caption{Cooling curves using the BSk21 EOS in steps of $0.01M_\odot$ with $\tilde{\lambda} = 10^3$ (Left panel),
  $\tilde{\lambda} = 10^4$ (Middle panel), and $\tilde{\lambda} = 10^5$ (Right panel), respectively.}
\label{fig:fig18}
\end{center}
\begin{center}
\includegraphics[width=1\linewidth]{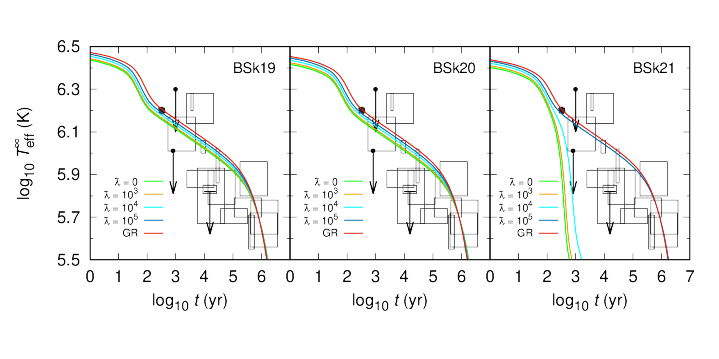}
\caption{Cooling curves with the different values of $\tilde\lambda$ and the EOS BSk19 (Left panel),
  the BSk20 (Middle panel), and the BSk21 (Right panel). Here, the mass is fixed as $M_s = 1.4~M_{\odot}$.
  The curve of GR in each panel corresponds to the infinity of $\tilde\lambda$ according to \cite{Kase:2019dqc}.}
\label{fig:fig18.1}
\end{center}
\end{figure}

Let us now discuss thermal evolution of INS in the scalar-tensor theory. Figs.~\ref{fig:fig15}$\sim$\ref{fig:fig17} present the cooling curves with various mass for the EOSs, BSk19, BSk20, and BSk21, respectively.
As expected from $Y_p$ distribution in Fig.~\ref{fig:fig10}, 
we find that the DU process occurs with the BSk21 EOS, while it does not occur with the other EOSs.
This is because the slope parameter $L$ is large enough only in the BSk21 EOS to derive the DU process.
  To see the cooling effect with the BSk21 EOS in more detail, Fig.~\ref{fig:fig18} shows the cooling
  curves with fixing the mass
in steps of $0.01M_{\odot}$. This figure shows that the threshold mass that the DU process begins to occur, $M_{\rm DU}$,
tends to be small as $\tilde{\lambda}$ decrease.
We find the threshold mass is in the range $1.25 < M_{\rm DU} / M_{\odot} \le 1.26$ for $\tilde{\lambda} = 10^3$,
$1.36 < M_{\rm DU} / M_{\odot} \le 1.37$ for $\tilde{\lambda} = 10^4$,
and $1.46 < M_{\rm DU} / M_{\odot} \le 1.47$ for $\tilde{\lambda} = 10^5$.
Hence, the DU process occurs even for smaller mass with smaller $\tilde \lambda$,
where the fifth force effect is more active.
Furthermore, this figure also indicates that the occurrence of the DU process is determined by the threshold mass
in a very narrow range smaller than $0.01 M_{\odot}$, which agrees with Ref.~\cite{Lim2017}.
The threshold value of $M_{\rm DU}$ obtained from the numerical results is slightly higher than $M_{\rm DU}$
expected from $Y_p$ distribution in Fig.~\ref{fig:fig10} and the mass-central density relation in Fig.~\ref{Mrho}.
This is explained by the fact that the occurrence of the DU process is not determined by the 
value of $Y_p$ at the center of a NS but the value of $Y_p$ at some finite radius. 

As one can see from Figs. \ref{fig:fig15} and \ref{fig:fig16} with the BSk19 and BSk20 EOSs, respectively, the effect of the fifth force looks small because the DU process does not occur for any mass.
Comparing the theoretical curves and the observational data of Cassiopeia A, we see that
the theoretical cooling curves depend on the value of $\tilde{\lambda}$. 
This is because the central density is lower as $\tilde{\lambda}$ is larger.
Therefore the effective region of the neutrino emissions becomes narrower. 
On the other hand, the cooling curves at the photon cooling stage can be
basically regarded as independent of neutrino cooling. This explains that
modification of gravity does not affect the cooling curves at the late stage $t\gtrsim 10^5~{\rm yr}$.

In order to compare the dependence of $\tilde\lambda$ on the cooling curves, Fig.~\ref{fig:fig18.1} shows
the cooling curves with fixing the mass $M_s = 1.4~M_{\odot}$ for the three EOSs.
One can check that the curves approach to the curve of GR as $\tilde{\lambda}$ increases.
Furthermore, one can see that the theoretical curves with the small value $\tilde\lambda$
fail to account for the observational data of Cassiopeia A. This feature is clear in the model with the EOS BSk21 because the DU process occurs. This figure exemplifies how the modification of gravity changes the cooling curves. 
By comparing the curve with $\tilde\lambda=0$ and the curve of the general relativity in Fig.~\ref{fig:fig17},
the change of $M_{\rm DU}$ is at most $0.4~M_{\odot}$ with the BSk21 EOS.
This panel shows that the cooling of NSs can be a tool for testing the scalar-tensor theories in the strong
regime. However, uncertainties of the modeling of NSs, e.g., the parameters of mass and the EOS,  must be
well fixed.

\section{Cooling Curves with Nucleon Superfluidity}

\begin{figure}[t]
\vspace*{-2.0cm}
\begin{center}
\includegraphics[width=\linewidth]{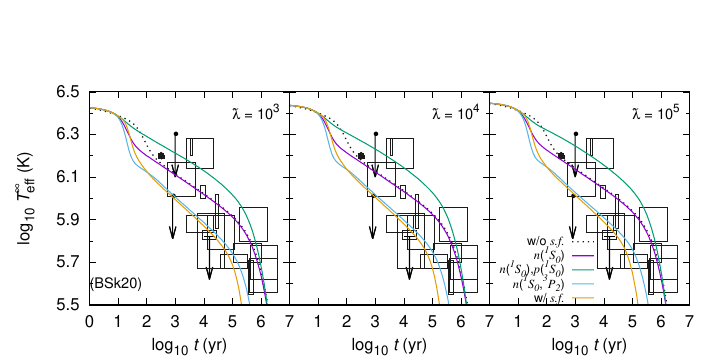}
\includegraphics[width=\linewidth]{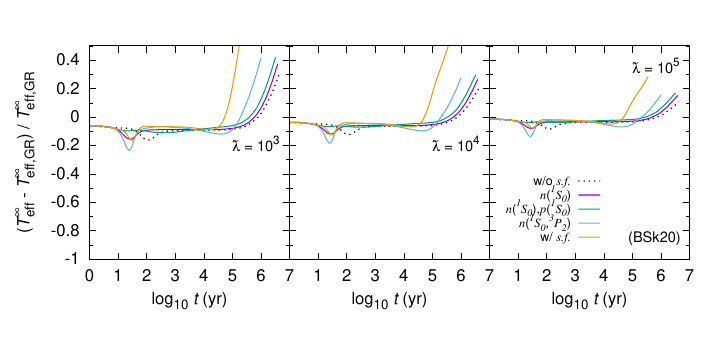}
\vspace*{-1.0cm}
\caption{Cooling curves (Top panel) and the derivation of surface temperature from that with GR (Bottom panel) with the BSk20 EOS, assuming $1.4 M_{\odot}$ stars. $\tilde{\lambda}=10^3$(Left panel), $10^4$(Middle panel), and $10^5$(Right panel). Dotted curves do not include superfluidity while the yellow curves all kinds of neutron(${}^1S_0,{}^3P_2$) and proton(${}^1S_0$) superfluidity. Except these cases, some of nucleon superfluidity are considered; curves with ``$n({}^1S_0,{}^3P_2)$" do not consider proton conductivity, for example.}
\label{fig:bsk20sf}
\end{center}
\end{figure}

\begin{figure}[t]
\vspace*{-0.0cm}
\begin{center}
\includegraphics[width=\linewidth]{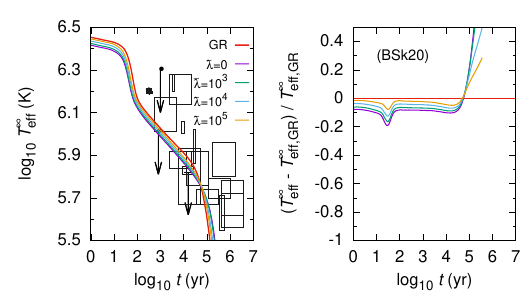}
\caption{Cooling curves (Left panel) and the derivation of surface temperature from that with GR (Right panel) with the BSk20 EOS, assuming $1.4 M_{\odot}$ stars. All kinds of neutron(${}^1S_0,{}^3P_2$) and proton(${}^1S_0$) superfluidity are considered. The difference of color denotes that of $\tilde{\lambda}$ values.}
\label{fig:bsk20sfg}
\end{center}
\end{figure}

\begin{figure}[htbp]
\vspace*{-2.0cm}
\begin{center}
\includegraphics[width=\linewidth,keepaspectratio,clip]{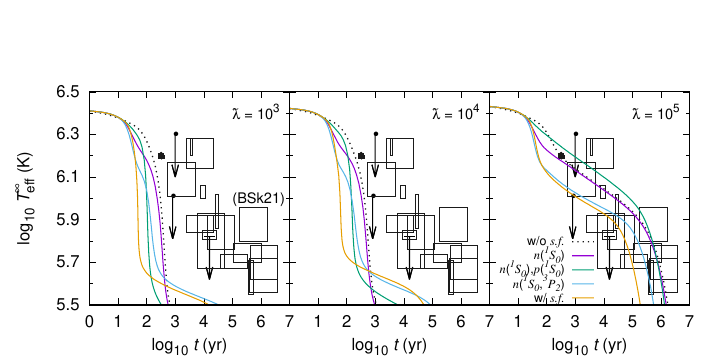}
\includegraphics[width=\linewidth,keepaspectratio,clip]{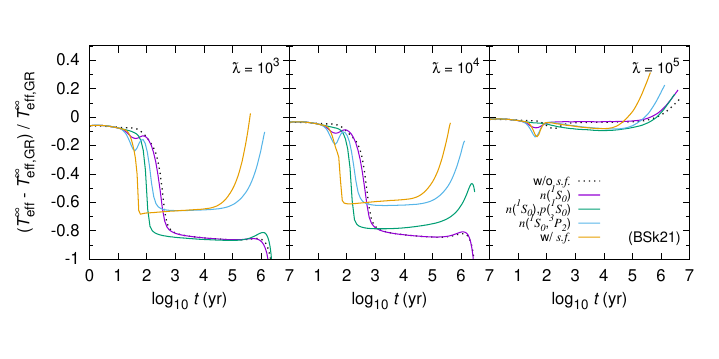}
\vspace*{-1.0cm}
\caption{The same as Fig.~\ref{fig:bsk20sf}, but with the BSk21 EOS.}
\label{fig:bsk21sf}
\end{center}
\end{figure}

\begin{figure}[h]
\vspace*{-0.0cm}
\begin{center}
\includegraphics[width=\linewidth]{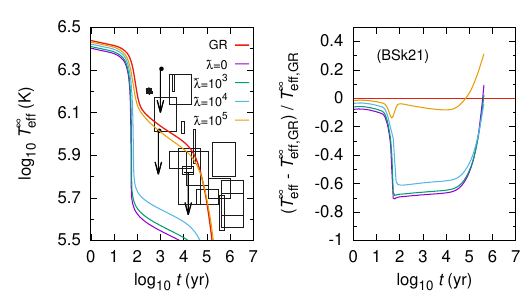}
\caption{The same as Fig.~\ref{fig:bsk20sfg}, but with the BSk21 EOS.}
\label{fig:bsk21sfg}
\end{center}
\end{figure}

The nucleon superfluidity, which is well known to be one of the important physics in the NS cooling theory was not taken into account in the results of the previous section. For completeness, we here calculate cooling curves with nucleon superfluidity based on the models described in the Sec.~\ref{coolingsec}. Throughout this section, we assume the mass of NSs to be $1.4~M_{\odot}$. We present the results with different superfluid models in Figs.~\ref{fig:bsk20sf} and \ref{fig:bsk21sf} which focus on the dependence of nucleon superfluid models. Moreover, we also show the $\tilde{\lambda}$ dependence of realistic cooling curves in which all kinds of nucleon superfluidity are considered in Figs.~\ref{fig:bsk20sfg} and \ref{fig:bsk21sfg}. Since the cooling behaviors with the BSk19 and BSk20 EOSs are similar as we see in the previous section, we do not show the case of BSk19 EOS.

 First of all, let us see the case of BSk20 EOS (See Fig.~\ref{fig:bsk20sf}), where the DU process is prohibited. Without nucleon superfluidity, the surface temperature gets lower around $t\sim10^2~{\rm yr}$ because the compactness of NSs due to the fifth force effect makes the neutrino luminosity higher. Such a difference from the GR case tends to appear as a {\it knee}, when the cooling rate are highly changed in a moment. Meanwhile, the surface temperature at photon cooling stage, whose timescale is approximately proportional to $M_{\rm NS}^{1/3}$ where $M_{\rm NS}$ is NS mass~(e.g., see Eq. (11.8.12) in Ref.~\cite{Shapiro1983}), is increased. When photon emission is dominant for INS cooling, low mass stars cool faster at photon cooling stage. This is why the photon luminosity decreases as $\lambda$ becomes smaller.  Even if we take nucleon superfluid effect, the qualitative tendency is not changed, but the variation of surface temperature is changed as in the Fig.~\ref{fig:bsk20sfg}. At neutrino cooling stage, the neutrons ${}^1S_0$ superfluidity shortens the timescale when the {\it knee} appears, and the change from the GR case finally appears around $t\sim50~{\rm yr}$ in this case. In addition, protons ${}^1S_0$ conductivity suppresses neutrino cooling while neutrons ${}^3P_2$ superfluidity promotes, and finally the former makes the difference from the GR case smaller while the latter makes larger with the age when the {\it knee} appears. Thus, surface temperature near the {\it knee} is slightly different between GR and scalar-tensor theory. This signature due to the fifth force effect might enable us to discriminate between GR and the scalar-tensor theory if a rapid cooling at $t\sim10^{2}~{\rm yr}$ is observed. At photon cooling stage, on the other hand, surface temperature is increased with scalar-tensor theory because the specific heat is reduced due to the pairing effect.

For the $1.4~M_{\odot}$ stars with the BSk21 EOS, the DU process occurs with $\tilde{\lambda}\lesssim10^4$ as we see in Fig.~\ref{fig:fig18.1}. In such models with scalar-tensor theory, the surface temperature is lower by $60\sim80\%$ for $t\gtrsim 10^2~{\rm yr}$ compared with the GR case (see left- and middle- bottom panels of Fig.~\ref{fig:bsk20sf}). If the nucleon superfluidity is considered, the cooling curves take values in higher-temperature regions because of the cooling suppression effect. Especially, the suppression effect of ${}^3P_2$ neutrons superfluidity is larger than that of other superfluidity in Fig.~\ref{fig:bsk21sf}. However, such a suppression effect is not large enough to cancel the effect of the DU process at least within the models adopted here.  Actually, the cooling curves without superfluidity (right panel in Fig.~\ref{fig:fig18.1}) and with superfluidity (left panel in Fig.~\ref{fig:bsk21sfg}) are not significantly different. Furthermore, the change of the surface temperature between the GR and the scalar-tensor theory within  minimal cooling scenario is at most 20\% (see right panels of Fig.~\ref{fig:bsk20sfg} and Fig.~\ref{fig:bsk21sfg} with $\tilde{\lambda}\gtrsim10^5$). This implies that the superfluid effect is not so important when the fast cooling processes occur. Although the effect of the cooling suppression due to the superfluid depends on the models, the actual threshold mass of the DU process is useful to distinguish between the theories of gravity even with nucleon superfluidity as suggested in previous section. 


Summarizing the above, the effect of fifth force basically tends to make the surface temperature around the {\it knee} highly decreased with neutrons superfluidity though the difference is not as large as the case of the DU process. Meanwhile, proton conductivity makes the difference of cooling curves from the GR case smaller. At the photon cooling stage, the surface temperature under scalar-tensor theory is increased. This tendency is similar with the slow cooling scenario. Therefore, 
we might distinguish between the 
GR and the scalar-tensor theory from 
the behavior of the {\it knee}, 
although the dominant cooling effect 
comes from the DU process depending 
on the models.
Moreover, if there are cold or rapidly cooled observations in the moment, such as the Cassiopeia A~\cite{Heinke2010}, it would be possible to put a constrain on the scalar-tensor theory.

\section{Discussion \& Conclusion}
\label{consec}

\begin{figure}[t]
\begin{center}
\includegraphics[width=0.6\linewidth]{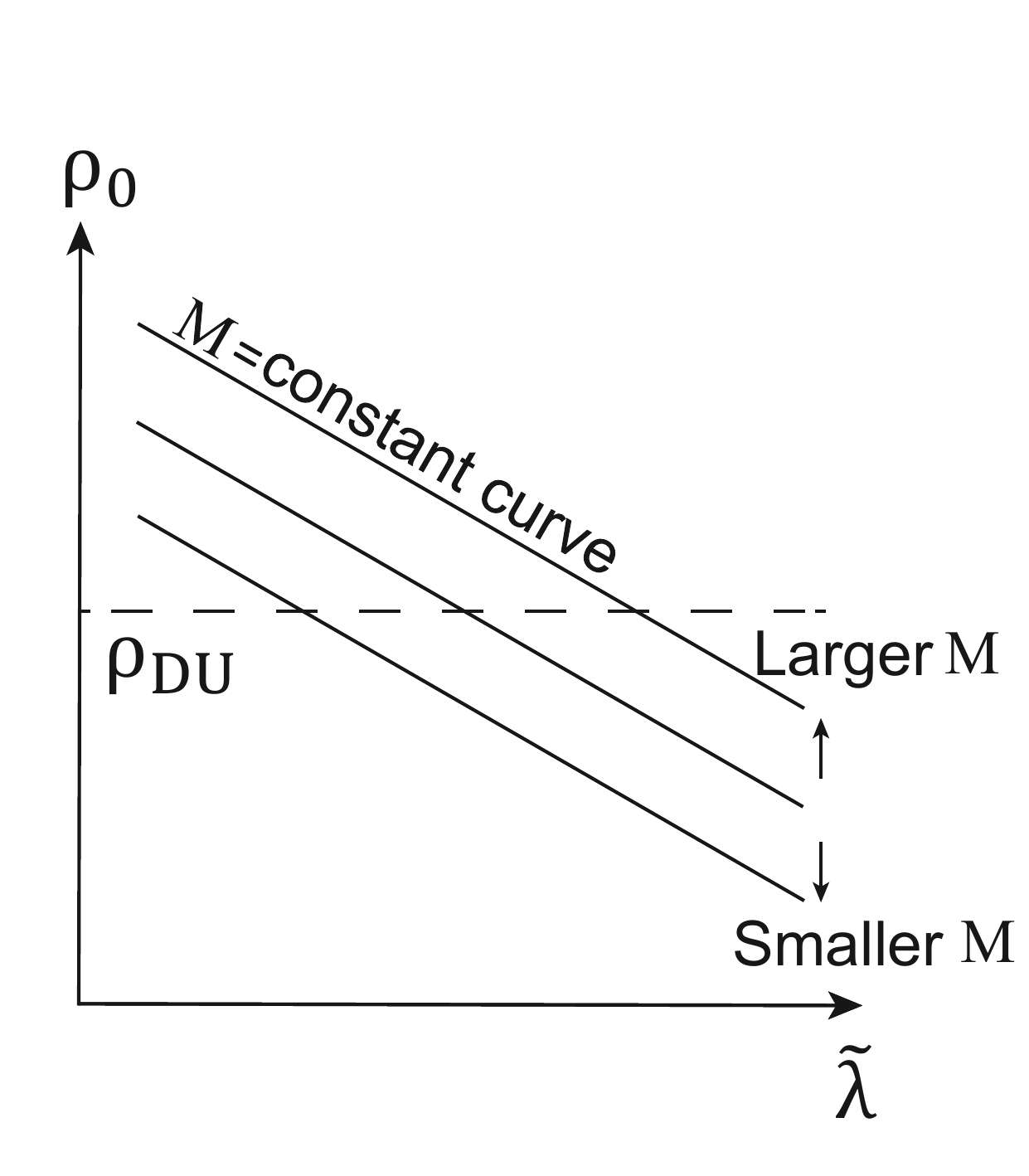}
\caption{ Schematic diagram to explain the dependence on the parameters  $\tilde{\lambda}$, $\rho_0$ and the mass $M$.
  $\rho_{\rm DU}$ means the threshold density above which the DU process occurs.}
\label{fig:schematic}
\end{center}
\end{figure}

We have considered the cooling effect of NSs in the scalar-tensor theory with
$\phi^4$ potential as well as massless Brans-Dicke theory discussed in Appendix~\ref{app:BD},
adopting BSk19, BSk20, and BSk21 as the EOS for NSs. In the scalar-tensor theory,
an additional scalar degree of freedom introduces so-called fifth force and modifies
the gravitational law predicted by general relativity.
If such fifth force is present, the structure of NSs can also differ from \textcolor{red}{the} one predicted in general relativity.
We first took a closer look at a fifth force effect in NSs and discussed how it modifies, e.g.,
the mass-radius relation.
We have confirmed that a screening of the fifth force is sufficiently active at high-density region, i.e.,
near the center of NSs. On the other hand, the fifth force effect cannot be still neglected near the surface
of NS even for large $\tilde \lambda$. Such an incomplete screening near the surface significantly affects
the mass-radius relation. Since the scalar field always invokes an attractive force in our case,
the net gravitational force is effectively stronger and NSs tend to be compact compared with general relativity,
as found for the cluster's gas density profile in Refs.~\cite{terukina1,terukina2}.
This implies that, if we fix the central density, the smaller $\tilde \lambda$ case has smaller mass as
confirmed in Fig.~\ref{Mrho}, and one can draw the constant mass curve in $\tilde \lambda$-$\rho_0$
plane\footnote{Strictly speaking, the constant mass curve is not exactly linear as drawn in Fig.~\ref{fig:schematic}.
However, it is enough to see the behavior of the cooling effect.} as in Fig.~\ref{fig:schematic}.  

For the nucleon superfluid effect on cooling curves, the behavior of the {\it knee}, where the cooling rate is drastically changed, is affected by the fifth force.  Hence, the properties of the {\it knee} can distinguish between the GR and the scalar-tensor theory, although the clear distinction between them would not be easy due to many model parameters such as EOSs, superfluidity, surface compositions. 
Observations of cold NSs or rapid-cooling NSs like Cassiopeia A might test the theories of gravity.  The observations of Cassiopeia provide us with
a precise cooling rate, which would be useful 
to test the scalar-tensor theory. 
This is left for a future work. 

Now let us discuss the cooling effect. Since the rapid cooling of NSs by the DU process does not occur for the EOSs,
BSk19 and BSk20, as discussed in Sec.~\ref{sec:EOS}, we hereafter focus on the case for BSk21.
In this case, whether the DU process occurs or not is solely determined by the central density of NSs as
shown in Fig.~\ref{fig:fig10}.
Thus, if the central density of NSs is above a certain value denoted as a threshold density $\rho_{\rm DU}$
(dashed line in Fig.~\ref{fig:schematic}), the DU process can be active in NS with the masses above the
corresponding
$M_{\rm DU}$ for each parameter $\tilde\lambda$. As clearly seen from Fig.~\ref{fig:schematic}, compared with
general relativity, even smaller-mass NSs can rapidly cool by the DU process in the scalar-tensor theory,
and $M_{\rm DU}$ tends to be small as $\tilde\lambda$ decreases, that is, as the fifth force effect is enhanced. 
The inverse of the parameter $\tilde\lambda$ typically represents the strength of the fifth force. Therefore,
we stress that these arguments on the rapid cooling of NSs can be also applied to other modified gravity theories
as long as a fifth force has an attractive nature\footnote{In contrast to scalar-tensor theories, the fifth force
  can be either attractive or repulsive in vector-tensor theories. Regarding the NS solutions in such theories,
  see Refs.~\cite{Chagoya:2017fyl,Kase:2017egk,Annulli:2019fzq,Kase:2020yhw,Ramazanoglu:2017xbl}, for example.}.
On the other hand, scalar-tensor theories that possess the Vainshtein mechanism might be 
difficult to see differences in a cooling curve since the interior structure of NSs is almost the
same as one in general relativity as reported in \cite{Ogawa2020}. Thus the information about masses
and cooling curves of NSs could provide a signature of the Chameleon force. It would be also interesting
to investigate cooling effects in other modified gravity theories having the symmetron mechanism
\cite{Hinterbichler:2010es} and the spontaneous scalarization \cite{Morisaki:2017nit}, and we leave these for future work.  

In this work, we assumed the specific scalar-tensor gravity models with the non-minimal coupling which mediates the fifth force as $F(\phi)=e^{2\phi/\left(\sqrt{6}M_{\rm pl}\right)}$ and the self-scalar coupling potential as $V(\phi)=\frac{1}{4}\lambda\phi^4$. If we consider other kinds of them, the structure and cooling of NSs naturally differ from those with the current model. However, the property to make NSs compact by the breaking down of the screening mechanism should be universal for the scalar-tensor models because the mediated fifth force is attractive. This implies some universal NS properties for gravity models; If the screening mechanism effectively works, the maximum mass and $M_{\rm DU}$ are higher, and the effect of nucleon superfluidity on cooling curves is lower. Hence, this work will provide a basis for these forthcoming discussions because the qualitative behaviors of NS structure and cooling curves are not affected by changing the law of gravity. 

We fixed the dimensionless self-coupling constant as $\tilde{\lambda}=0,10^3,10^4,10^5$ in this study.
Then, the current models are inconsistent with the E\"{o}t-Wash torsion balance experiment demanding $\lambda\gtrsim\mathcal{O}(1)$ or $\tilde{\lambda}\gtrsim\mathcal{O}(10^{78})$ (e.g., \cite{Burrage2016}). Since the screening mechanism effectively works with $\tilde{\lambda}=10^5$. the NS structure and cooling curves should be almost the same as those with GR in the experimentally good $\lambda$ value. For calculation with such an extremely large $\lambda$, however, there is numerical difficulty as presented in \cite{Kase:2019dqc} (such problems are mentioned in other literatures as well). By tackling this numerical problem, we hope that the above point will be investigated in detail as a future work.

\section*{Acknowledgment}

We would like to thank T. Noda, H. Sotani, and S. Nagataki for fruitful discussions.
We also thank M. Machida and K. Nakazato for the communications 
related to the topic of the present paper. 
R. Kase is supported by the Grant-in-Aid for Young Scientists B of the JSPS No.17K14297.
This work was supported in part by JSPS Grant-in-Aid for Scientific Research 
Nos.~JP17K14276 (R. Kimura).
KY is supported by MEXT/JSPS KAKENHI Grant No. 15H05895, No. 16H03977 No. 17K05444, No. 17H06359.

\appendix

\section{Einstein frame description}
\label{app:EF}
In this appendix, we summarize the relation between the Jordan frame and the Einstein frame. 
The Einstein frame description is useful to see a screening of the fifth force effect and provides further insight of a dynamics of the scalar field.
The action with a non-minimal coupling of $\phi$ with gravity can be transformed into one with a minimal coupling through the following invertible mapping, so-called conformal transformation,
\ba
(g_{\mu \nu})_{\rE}=F(\phi) g_{\mu \nu}\,,
\label{conformal}
\ea
where the subscript E represents quantities defined in the Einstein frame throughout this paper, and $(g_{\mu \nu})_{\rE}$ is, for example, the metric in the Einstein frame.
Then, the action \eqref{action} in the Einstein frame can be written as 
\ba
{\cal S}_{\rE}=\int \rd^4 x \sqrt{-g_{\rE}} \left[ 
\frac{M_{\rm pl}^2}{2} R_{\rE} -\frac{1}{2} 
g^{\mu \nu}_{\rE} \partial_{\mu} \phi 
\partial_{\nu} \phi-V_{\rE}(\phi) \right]
+\int \rd^4 x\,{\cal L}_m \left( F^{-1}(\phi) 
(g_{\mu \nu})_{\rE}, \Psi_m \right)\,,
\label{actionE}
\ea
where the potential for the scalar field in the Einstein is given by 
\ba
V_{\rE} (\phi)=\frac{V(\phi)}{F^2(\phi)}\,.
\ea
In this frame, although the matter field directly couples to the scalar field through the relation \eqref{conformal}, the gravity sector is the one in general relativity. We define the energy-momentum tensor for the matter field in the Einstein frame as
$(T_{\mu \nu})_{\rE}=-(2/\sqrt{-g_{\rE}}) \delta S_m/ 
\delta g^{\mu \nu}_{\rE}$, and then the Einstein equation and  the Euler-Lagrange equation for the scalar field are given by
\ba
\mpl^2 (G_{\mu\nu})_{\rm E} &=&
(T_{\mu\nu})_{\rm E}+ (T_{\mu\nu}^{(\phi)})_{\rm E}  \,,\\
\square_{\rE} \phi - \frac{\partial V^{\rm eff}_{\rm E}}{\partial \phi}
&=&0\,,
\label{phiEin}
\ea
where $(G_{\mu\nu})_{\rm E}$ and $\square_{\rE} $ are respectively the Einstein tensor and d'Alembert operator evaluated with the metric in the Einstein frame $(g_{\mu\nu})_{\rm E}$, the energy-momentum tensor for the scalar field is given by
\ba
(T_{\mu\nu}^{(\phi)})_{\rm E} =\nabla_\mu \phi \nabla_\nu \phi - g_{\mu\nu} \left({1\over 2}\nabla_\alpha \phi \nabla^\alpha \phi +V(\phi) \right)\,,
\ea
and $V^{\rm eff}_{\rm E}$ is the effective potential defined as
\ba
V^{\rm eff}_{\rm E} \equiv
V_{{\rE}}+\frac{Q}{M_{\rm pl}} 
\left( \rho_{\rE}-3P_{\rE} \right)\phi \,.
\label{Veff}
\ea
Since the matter field directly couples to the scalar field, the energy-momentum conservations for each components do not identically hold in this frame, and we instead have 
\ba
\nabla^\mu 
\left[(T_{\mu\nu}^{(\phi)})_{\rm E}  + (T_{\mu\nu})_{\rm E}\right]=0 \,.
\ea

As in the Jordan frame, 
we introduce the spherically symmetric and static 
background whose line element as 
${\rm d}s_{\rE}^2 = (g_{\mu\nu})_{\rm E} dx_{\rm E}^\mu dx_{\rm E}^\nu=-f_{\rE}(r_{\rE}) \rd t^2+h_{\rE}^{-1}(r_{\rE}) \rd r_{\rE}^2 
+r_{\rE}^2 ( \rd \theta^2+\sin^2 \theta\,\rd \varphi^2 )$,
and the energy density and pressure for the perfect fluid in the Einstein frame as 
$(T^{\mu}_{\nu})_{\rE}={\rm diag} (-\rho_{\rE}, 
P_{\rE}, P_{\rE}, P_{\rE})$. 
Then the Einstein equation gives two independent equations :
\ba
\frac{1}{ 4\pi r_{\rE}^2 }\frac{\rd {\cal M}_{\rE}}{\rd r_{\rE}}
&=& \rho_{\rE}
+\frac{h_{\rE}}{2} 
\left( \frac{\rd \phi}{\rd r_{\rE}} \right)^2+V_{\rE} \,,\\
M_{\rm pl}^2 \left[
\frac{h_{\rE}}{r_{\rE}f_{\rE}}\frac{\rd f_{\rE}}{\rd r_{\rE}}
+ \frac{h_{\rE}-1}{r_{\rE}^2}
\right]
&=&
P_{\rE} + 
{h_{\rE} \over 2} 
\left( \frac{\rd \phi}{\rd r_{\rE}} \right)^2 
-V_{\rE}\,,
\label{MEeq}
\ea
where the mass function in the Einstein frame is defined by $h_{\rE} (r_{\rE})=1-{2G{\cal M}_{\rE} (r_{\rE})}/{r_{\rE}}$. The continuity equation for the matter and the equation of motion for the scalar field are given by
\ba
\frac{\rd P_{\rE}}{\rd r_{\rE}}+\frac{1}{2f_{\rE}}
\frac{\rd f_{\rE}}{\rd r_{\rE}} \left( \rho_{\rE}+P_{\rE} 
\right)+\frac{Q}{M_{\rm pl}} \left( \rho_{\rE}-3P_{\rE} 
\right) \frac{\rd \phi}{\rd r_{\rE}}&=&0\,, \label{continuityE}\\
\frac{\rd^2 \phi}{\rd r_{\rE}^2}+\left[ \frac{2}{r_{\rE}}
+\frac{1}{2} \frac{\rd}{\rd r_{\rE}} \ln \left( f_{\rE} h_{\rE} 
\right) \right] 
\frac{\rd \phi}{\rd r_{\rE}}-\frac{1}{h_{\rE}} \left[ 
V_{\rE, \phi}+\frac{Q}{M_{\rm pl}} 
\left( \rho_{\rE}-3P_{\rE} \right) \right]&=&0\,.
\label{phiEeq}
\ea
Here the second term and third term in the continuity equation \eqref{continuityE} respectively represent the force from the Einstein frame metric and the scalar field. 
In analogy with the density parameters in the Friedmann equation, we introduce the following dimensionless parameters  
\ba
\Omega_\phi^{(h)} = \frac{(T^{0}_{~0}{}^{(\phi)})_{\rm E} }{(G^0_{~0})_{\rm E} }, \qquad 
\Omega_\phi^{(f)} = \frac{(T^{1}_{~1}{}^{(\phi)})_{\rm E} }{(G^1_{~1})_{\rm E} }\,,
\label{para1}
\ea
which measures the fraction of the energy-momentum tensor for the scalar field in the Einstein equation. 
In addition to these parameters, we also introduce the force fraction parameter 
\ba
E_\phi = 
\left|\left(\frac{\rd P_{\rE}}{\rd r_{\rE}}\right)^{-1}
\frac{Q}{M_{\rm pl}} \left( \rho_{\rE}-3P_{\rE} 
\right) \frac{\rd \phi}{\rd r_{\rE}}\right|\,.
\label{para2}
\ea
For $ \rho_{\rE}-3P_{\rE} >0$, the scalar field is an attractive force,  and this parameter then satisfies $0 \leq E_\phi \leq 1$. Since the derivative of the pressure vanishes at the origin and outside of NSs, we define this parameter within $0< r < r_s$.

Finally, by using the fact that ${\rm d}s_{\rE}^2=F\rd s^2$, we can relate quantities in the Jordan and Einstein frame as 
\ba
r &=&  e^{Q\phi/M_{\rm pl}} r_{\rE}\,,\\
f(r) &=& e^{2Q\phi/M_{\rm pl}} f_{\rE} (r_{\rE})\,,\\
h(r) &=& h_{\rE} (r_{\rE}) \left( 1+\frac{Qr_{\rE}}{M_{\rm pl}}
\frac{\rd \phi}{\rd r_{\rE}} \right)^2\,,\\
\rho_{\rE}&=&\frac{\rho}{F^2}\,,\\
P_{\rE}&=&\frac{P}{F^2}\,.
\label{hcore}
\ea
With these relation, one can compute the Einstein frame quantities from one in the Jordan frame obtained in numerical calculation.

\section{Chameleon mechanism}
\label{app:chameleon}
\begin{figure}[thbp]
	\begin{center}
		\includegraphics[scale=1.0]{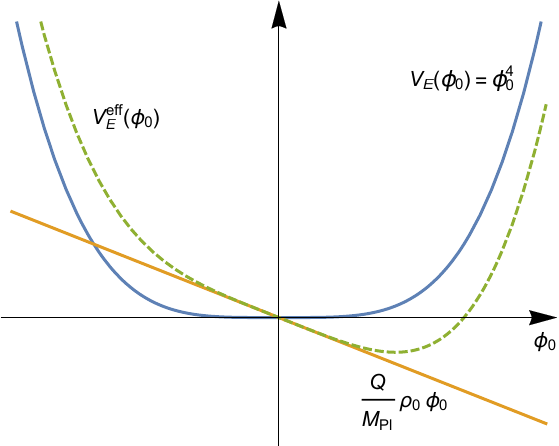}
	\end{center}
	\caption{\label{fig:Veff}
		Schematic figure of the effective potential $V_{\rm E}^{\rm eff}(\phi_0)$. The potential for the scalar field is chosen to be $V_{\rm E}(\phi_0) = \phi_0^4$ here.
	}
\end{figure}

In this appendix, we review the Chameleon mechanism in a non-relativistic setup (see e.g., \cite{Burrage:2017qrf} for the detail.). Hereafter, we work in the Einstein frame summarized in the appendix \ref{app:EF}. 
Let us first consider the geodesic equation in the Einstein frame,
\ba
{d^2 x_{\rm E}^\mu \over d\tau^2} + \Gamma_{\alpha\beta}^\mu {d x_{\rm E}^\alpha \over d\tau}{d x_{\rm E}^\beta \over d\tau}=0 \,,
\ea
where $\tau$ represents a proper time in the Einstein frame, and the Christoffel symbol $\Gamma^\mu_{\alpha\beta}$ is evaluated  by $g_{\mu\nu} = F^{-1} (\phi) (g_{\mu\nu})_{\rm E}$. Then taking the non-relativistic limit, we arrive at the Newtonian equation of motion \cite{Sakstein:2015oqa}, 
\ba
\ddot{x}^i_{\rm E} = - \partial^i_{\rm E} \Phi_N - \partial^i_{\rm E} \Phi_\phi\,,
\ea
where $\Phi_N$ is the Newtonian gravitational potential, and $\Phi_\phi$ is the scalar potential defined as $\Phi_\phi = Q \phi / \mpl$. Therefore, the fifth force due to the scalar field given by
\ba
{\bf F}_\phi = - {Q \over \mpl} \nabla_{\rm E} \phi \,.
\ea 
Now we would like to find a profile of the scalar field. To this end, we, for simplicity, consider a point source of mass $M$ in a medium of homogeneous background density $\rho_0$ and split the scalar field as the background value $\phi_0$ and perturbation $\delta\phi$.  Assuming the scalar field is located at the minimum of the effective potential \eqref{Veff},  the background value $\phi_0$ is then simply determined by 
\ba
{dV_{\rm E}^{\rm eff} \over d\phi} \Bigg|_{\phi = \phi_0} = 0 \,.
\ea
Now we consider the scalar field equation \eqref{phiEin} around this background, which is given by
\ba
\nabla_{\rm E}^2 \delta \phi - m_{\rm eff}^2(\phi_0) \delta \phi = {Q \rho \over \mpl} \,,
\ea
where $m_{\rm eff}$ is the effective mass defined as
\ba
m_{\rm eff}^2 (\phi) \equiv {d^2 V_{\rm E}^{\rm eff} (\phi)\over d\phi^2} \,, 
\ea
and the energy density of a point source is given by  $\rho = M \delta^3(x)$.
Then, the solution of this equation can be written as 
\ba
\Phi_\phi =
{Q \over \mpl} \delta \phi =2 Q^2  {GM \over r} e^{- m_{\rm eff} r} \,.
\ea
When $m_{\rm eff} r \ll 1$, the fifth force is comparable to the Newtonian one, $-\nabla \Phi_N$, if $Q \sim {\cal O} (1)$.  On the other hand, when $m_{\rm eff} r \gg1$, the scalar force is short ranged, and the fifth force can be negligible due to the exponential factor, that is, the fifth force is screened. Thus, whether the Chameleon screening works or not is determined by the Compton wavelength of the scalar field $\lambda_C = m_{\rm eff}^{-1}$. As one can see the schematic plot of the effective potential in Fig.~\ref{fig:Veff}, the shape of the effective potential depends on the density of environments. As the density $\rho_0$ increase, the effective mass $m_{\rm eff}$ becomes larger, and the Compton wavelength $\lambda_C$ becomes shorter. This tells us that the Chameleon screening becomes more effective in a high density environment. 
Note that the effective potential also depends on pressure in the case of neutron stars, and also the profile of the scalar field including relativistic effects becomes more complicated.

\section{Massless Brans-Dicke theory}
\label{app:BD}
\begin{figure}[thbp]
  \begin{center}
               \includegraphics[scale=0.25]{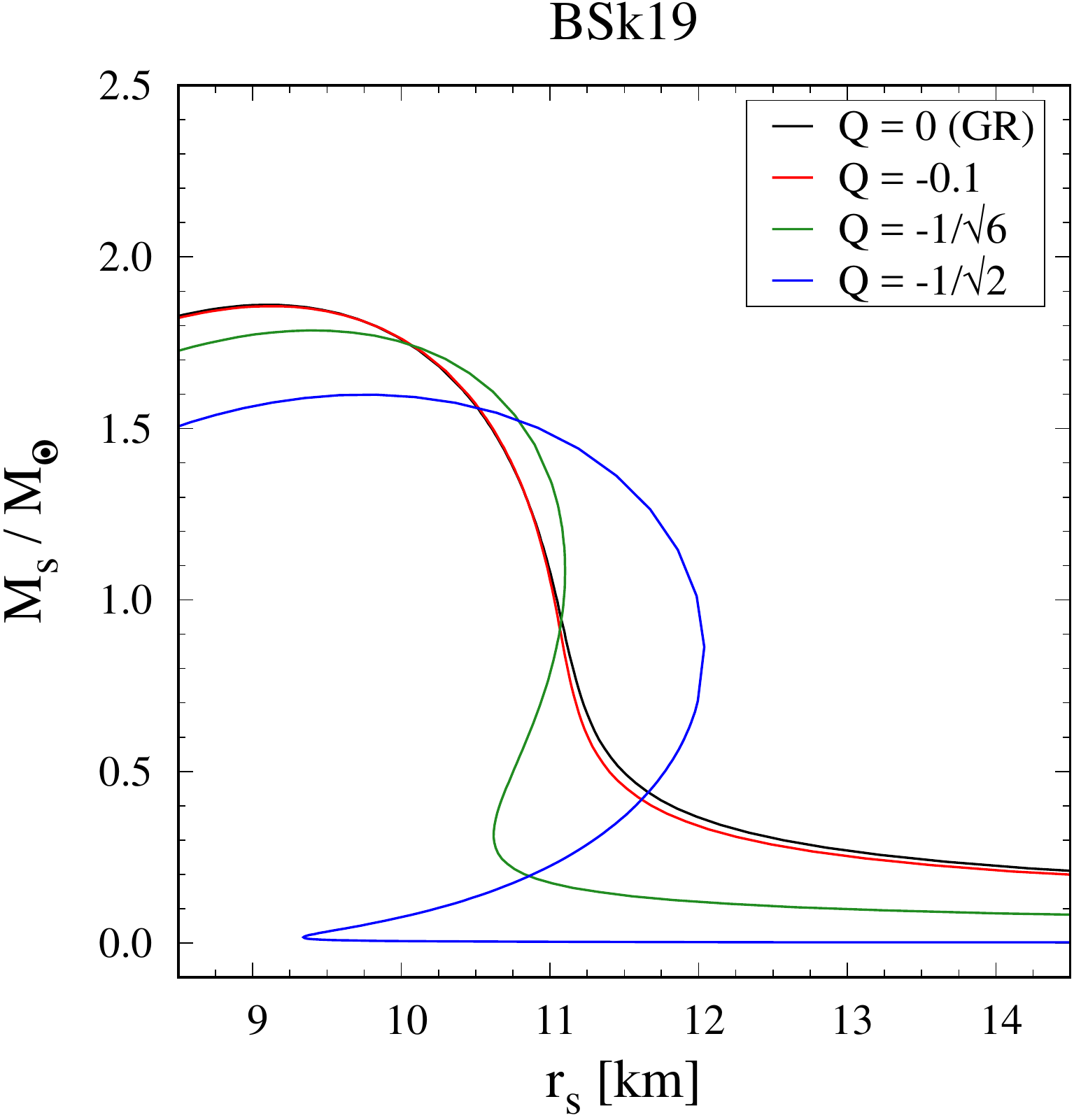}
                \includegraphics[scale=0.25]{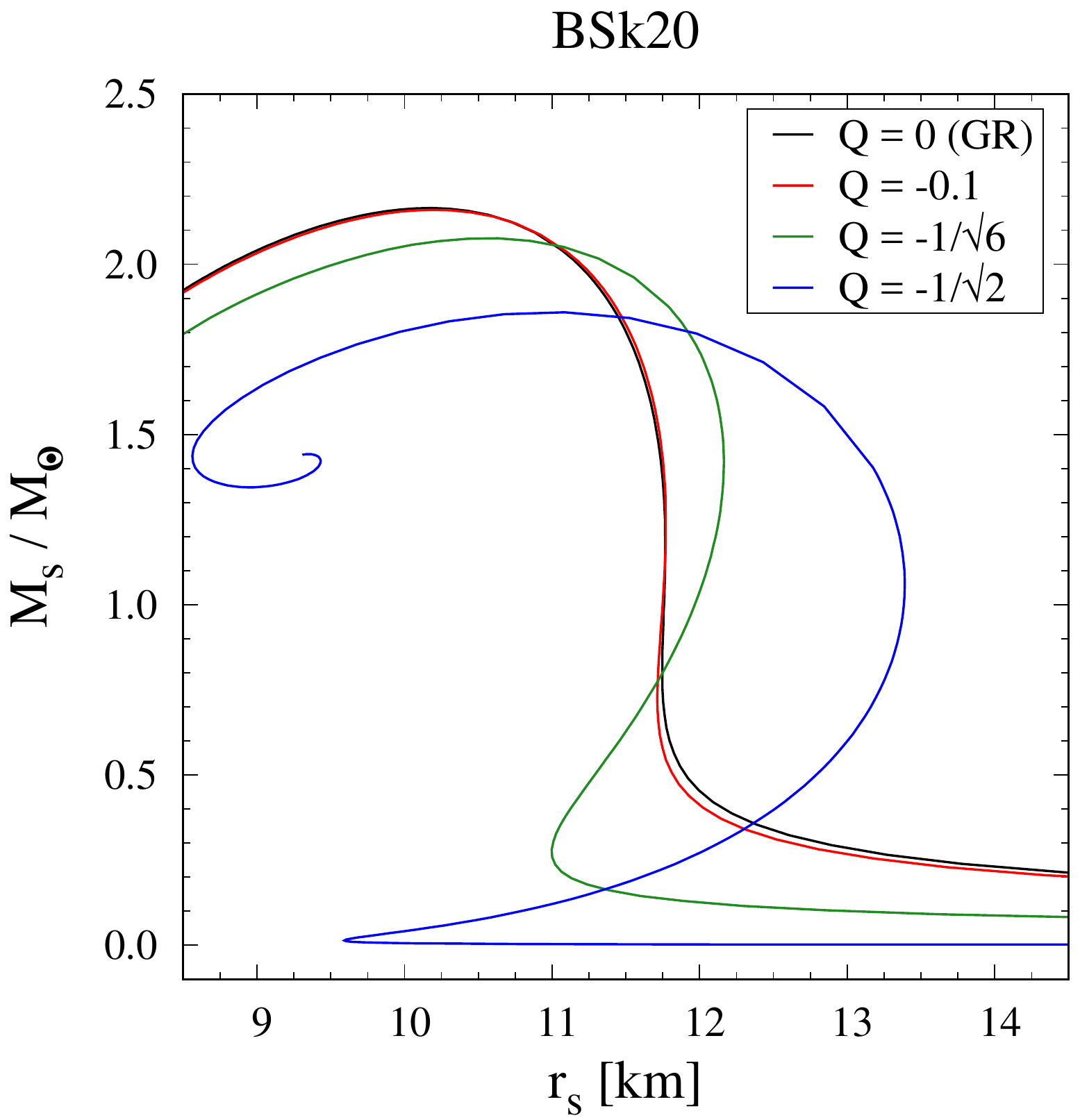}\\
                \includegraphics[scale=0.25]{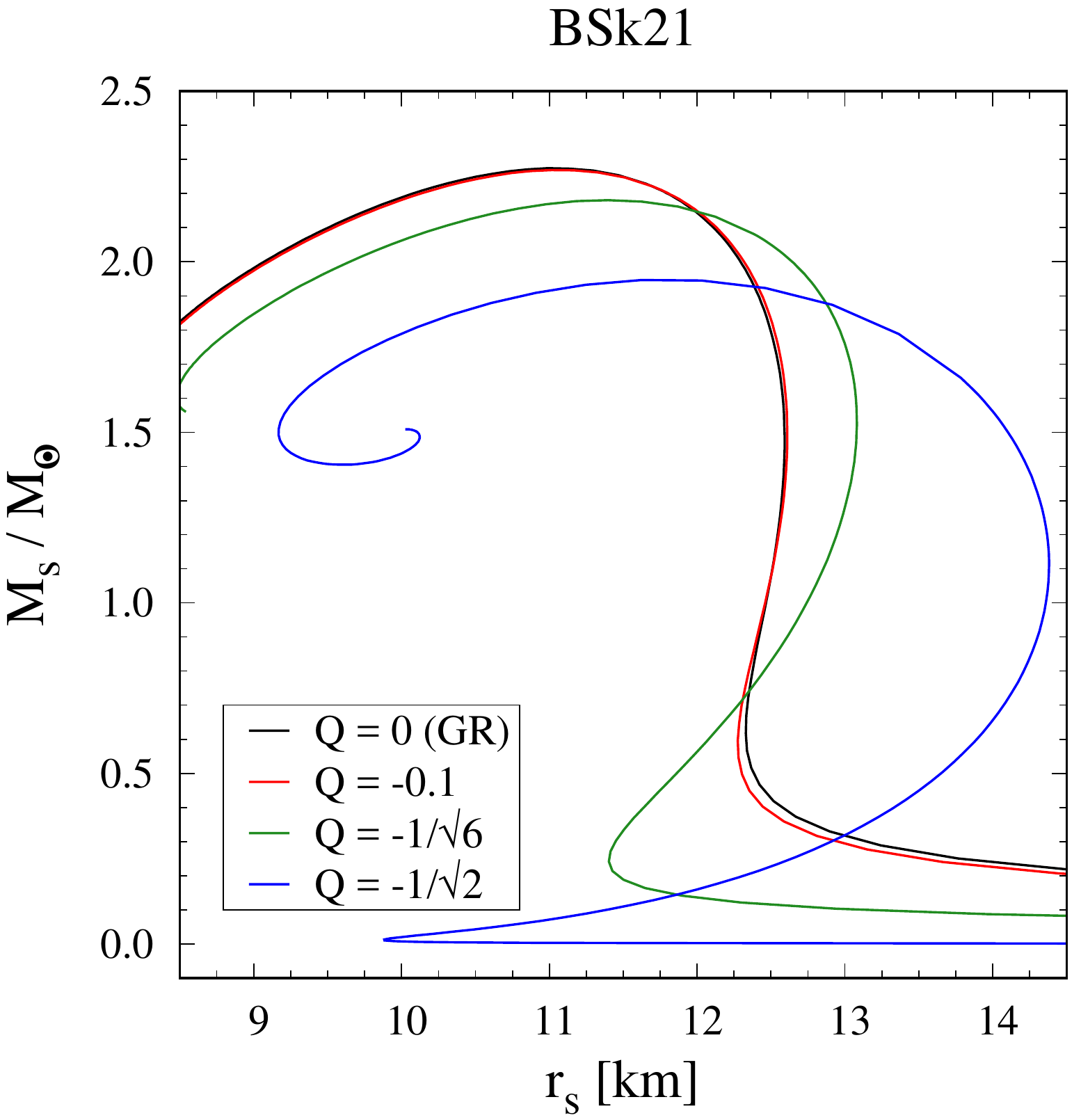}
  \end{center}
  \caption{\label{MRmassless}
		Mass-radius relation for BSk19 (upper left panel), BSk20 (upper right panel), and BSk19 (lower panel) EOSs in the massless BD model.
	}
\end{figure}

In this appendix, we consider the massless Brans-Dicke theory whose solutions of NSs are discussed in \cite{Kase:2019dqc}.
The massless Brans-Dicke theory is given by the action \eqref{action} with $V(\phi)=0$, and 
the relation between the parameter $Q$ and the original Brans-Dicke parameter is given by $2Q^2 = 1/(3+2\omega_{BD})$ \cite{Brans}. This model is indistinguishable from general relativity with the canonical scalar field in the limit of $Q \to 0$. Although this model for $Q \sim {\cal O}(1)$ is not completely consistent with the solar-system experiments
\cite{Will:2014kxa} due to a significant modification of gravitational law at all scales, we, for demonstration, consider the effect of NS cooling in this model. 
Although we adopt the different EOSs (BSk19, BSk20, and BSk21) from ones used in \cite{Kase:2019dqc}, the details of the structure of NSs are qualitatively similar here. Therefore, we only plot mass-radius relations for the EOSs, BSk19, BSk20, and BSk21 in Fig.~\ref{MRmassless}. 

\begin{figure}[thbp]
\begin{center}
\includegraphics[width=0.9\linewidth]{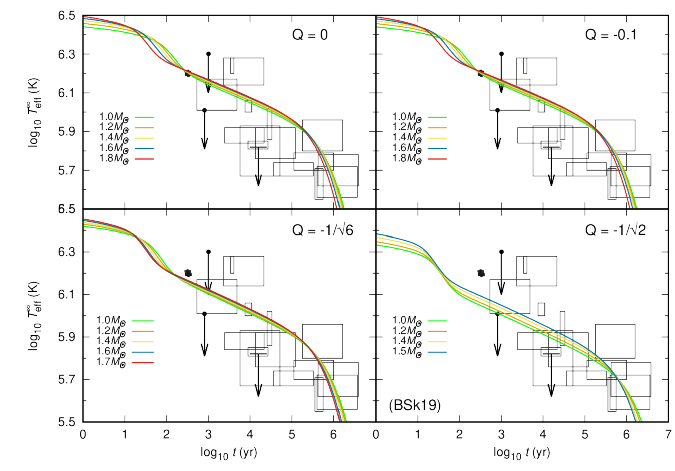}
\caption{Cooling curves with massless BD models using BSk19 EOS. }
\label{fig:fig12}
\vspace{1cm}
\includegraphics[width=0.9\linewidth]{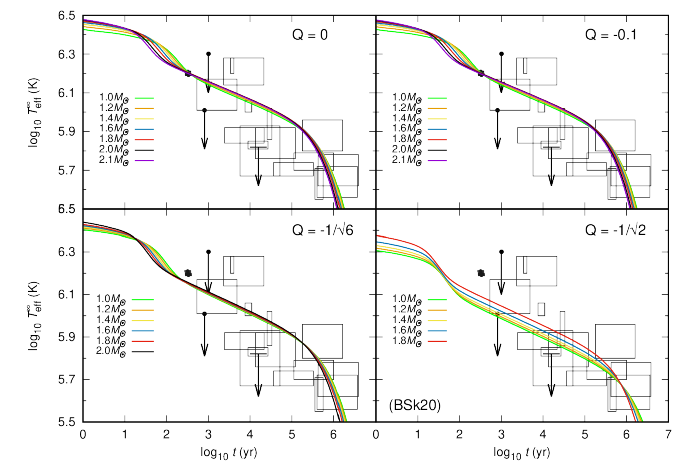}
\caption{Same as Fig.~\ref{fig:fig12}, but with the BSk20 EOS.}
\label{fig:fig13}
\end{center}
\end{figure}

Furthermore, we plot the cooling curves with massless Brans-Dicke models in Figs.~\ref{fig:fig12}$\sim$\ref{fig:fig15}. The change of the cooling curves under scalar-tensor theory can be explained in case of Figs.~\ref{fig:fig12}$\sim$\ref{fig:fig14.1}; With small $|Q|$ the central density with a fixed mass becomes higher. In BSk21 EOS, $M_{\rm DU}$ becomes drastically lower with higher--$|Q|$ and in case of $Q=-1/\sqrt{2}$ (massless dilaton), the cooling curves cannot explain all observations though some uncertain parameters such as superfluid effect are not considered. The effect of cooling curves by changing scalar-tensor theories through $|Q|$ can be qualitatively understood by replacing $\tilde{\lambda}$ with $1/|Q|$ in Fig.~\ref{fig:schematic}. These figures enable us to clearly see the positive correlation between $|Q|$ and the central density, related to $M_{\rm DU}$, from a fixed mass.

We have also calculated the cooling curves including the effect of nucleon superfluidity in the massless Brans-Dicke model as in Figs.~\ref{fig:lbsk20sfg} and \ref{fig:lbsk21sfg}.	
As in the case of 
the DU process, the $|Q|$ dependence on the cooling curves is qualitatively the same as the tendency of $1/\tilde{\lambda}$ in Figs.~\ref{fig:bsk20sfg} and \ref{fig:bsk21sfg}. In $Q=-0.10$, the cooling curves are not almost changed with the GR case, but in $Q=-1/\sqrt{6},-1/\sqrt{2}$, the cooling curves basically tend to locate in lower-temperature regions compared with the GR case. This is because the larger-$|Q|$ values have higher-central density, that is, stronger neutrino emissions, as with the case of cooling curves without superfluidity.

\vspace*{1cm}

\begin{figure}[thbp]
\begin{center}
\includegraphics[width=0.9\linewidth]{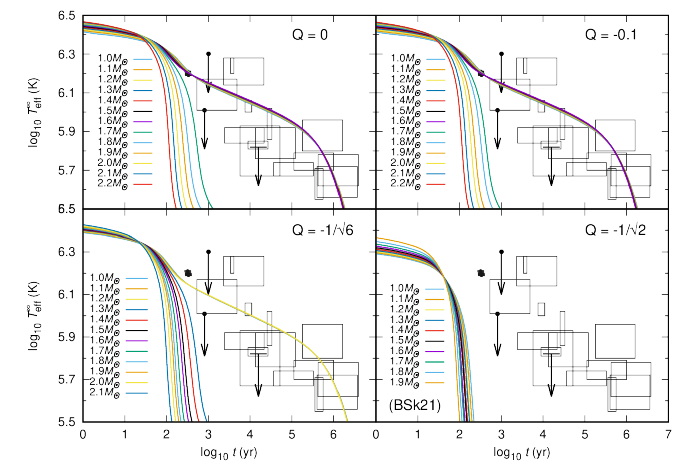}
\caption{Same as Fig.~\ref{fig:fig12}, but with the BSk21 EOS.}
\label{fig:fig14}
\end{center}
\end{figure}
\begin{figure}[thbp]
\begin{center}
\includegraphics[width=0.9\linewidth]{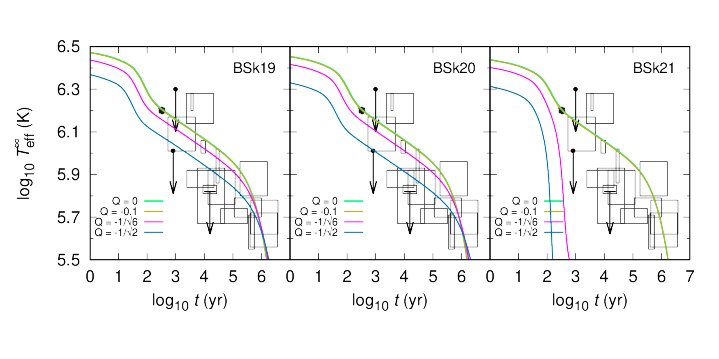}
\caption{Cooling curves with massless BD model with $M_s = 1.4~M_{\odot}$.}
\label{fig:fig14.1}
\end{center}
\end{figure}

\begin{figure}[htbp]
\vspace*{-1.0cm}
\begin{center}
\includegraphics[width=\linewidth]{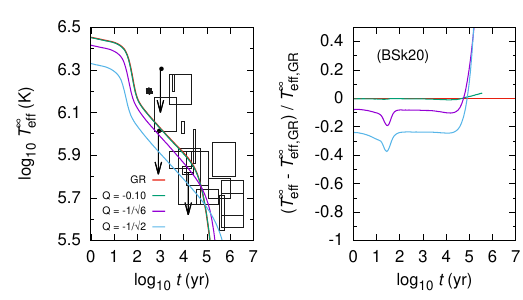}
\caption{Same as Fig.~\ref{fig:bsk20sfg}, but with mass-less Brans-Dicke theory.}
\label{fig:lbsk20sfg}
\end{center}
\end{figure}

\begin{figure}[htbp]
\vspace*{-1.0cm}
\begin{center}
\includegraphics[width=\linewidth]{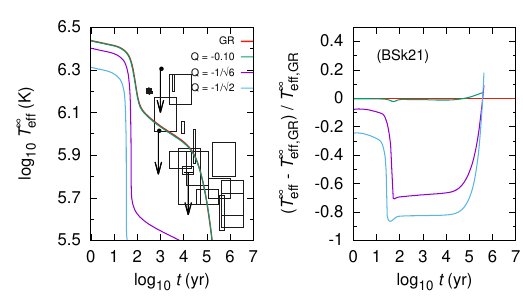}
\caption{Same as Fig.~\ref{fig:bsk21sfg}, but with mass-less Brans-Dicke theory.}
\label{fig:lbsk21sfg}
\end{center}
\end{figure}


\end{document}